\documentclass[apjl]{emulateapj}
\usepackage{graphicx,url}
\usepackage{color}

\def\rxte{{\it RXTE}}
\def\xmm{{\it XMM-Newton}}

\def\suz{{\it Suzaku}}
\def\swi{{\it Swift}}
\def\nustar{{\it NuSTAR}}
\def\xillver{{\tt xillver}}
\def\relxill{{\tt relxill}}

\def\pcacorr{{\sc pcacorr}}

\shorttitle{X-ray Reflection Spectroscopy of GX~339--4}
\shortauthors{Garc\'{\i}a \& et al.}

\begin{document}



\title{X-ray Reflection Spectroscopy of the Black Hole GX~339--4: 
Exploring the Hard State with Unprecedented Sensitivity}

\author{Javier~A.~Garc\'ia\altaffilmark{1}, James~F.~Steiner\altaffilmark{1},
  Jeffrey~E.~McClintock\altaffilmark{1}, Ronald~A.~Remillard\altaffilmark{2},
  Victoria~Grinberg\altaffilmark{2}, Thomas~Dauser\altaffilmark{3}}

\altaffiltext{1}{Harvard-Smithsonian Center for Astrophysics,
  60 Garden St., Cambridge, MA 02138 USA; javier@head.cfa.harvard.edu,
  jem@cfa.harvard.edu, jsteiner@head.cfa.harvard.edu}

\altaffiltext{2}{MIT Kavli Institute for Astrophysics and Space Research,
  MIT, 70 Vassar Street, Cambridge, MA 02139, USA; rr@space.mit.edu,
  grinberg@space.mit.edu}

\altaffiltext{3}{Dr.\ Karl Remeis-Observatory and Erlangen Centre for 
  Astroparticle Physics, Sternwartstr.~7, 96049 Bamberg, Germany;
  thomas.dauser@sternwarte.uni-erlangen.de}

%

\begin{abstract}

  We analyze {\it simultaneously} six composite {\it RXTE} spectra of
  GX~339--4 in the hard state comprising 77 million counts collected
  over 196 ks.  The source spectra are ordered by luminosity and span
  the range 1.6\% to 17\% of the Eddington luminosity.  Crucially, using
  our new tool {\tt pcacorr}, we re-calibrate the data to a precision of
  0.1\%, an order of magnitude improvement over all earlier work.  Using
  our advanced reflection model {\tt relxill}, we target the strong
  features in the component of emission reflected from the disk, namely,
  the relativistically-broadened Fe K emission line, the Fe K edge and
  the Compton hump.  We report results for two joint fits to the six
  spectra: For the first fit, we fix the spin parameter to its maximal
  value ($a_*=0.998$) and allow the inner disk radius $R_{\rm in}$ to
  vary.  Results include (i) precise measurements of $R_{\rm in}$, with
  evidence that the disk becomes slightly truncated at a few percent of
  Eddington; and (ii) an order-of-magnitude swing with luminosity in the
  high energy cutoff, which reaches $>890$~keV at our lowest luminosity.
  For the second fit, we make the standard assumption in estimating spin
  that the inner edge of the accretion disk is located at the innermost
  stable circular orbit ($R_\mathrm{in} = R_\mathrm{ISCO}$) and find
  $a_* = 0.95^{+0.03}_{-0.05}$ (90\% confidence, statistical).  For both
  fits, and at the same level of statistical confidence, we estimate
  that the disk inclination is $i = 48\pm 1$~deg and that the Fe
  abundance is super-solar, $A_\mathrm{Fe} = 5\pm1$.

\end{abstract}


%
%
%
%
\section{Introduction}\label{sec:intro}

GX~339--4 is one of the most thoroughly studied of the roughly 50 known
black-hole X-ray binaries. Its orbital period is around 1.7~days, and for the
best candidate period of 1.7557~days the mass function is $5.8\pm0.5~M_{\odot}$
\citep{hyn03}. Like nearly all black hole binaries, the X-ray source is
transient, having undergone more than a dozen outburst cycles since its
discovery in the early 1970s by \cite{mar73}. During a cycle, GX339--4 often
exhibits all known X-ray states, which unfold in the canonical pattern
\citep{rem06,dun10}. During the rising phase, the source can reach exceptional
luminosities in the hard state, which is the focus of this paper.
  
The hard state is strongly dominated by a hard power-law component ($\Gamma
\sim 1.6$). The thermal component, which contributes $\lesssim20$\% of the
2--20~keV flux, is faint and cool ($kT\lesssim0.2$~keV) compared to the thermal
state \citep{rem06}. The Fe K line is a ubiquitous spectral feature. Strong
variability is a hallmark of the hard state (rms power $>10$\% in the band
0.1--10~Hz), while QPOs may be either present or absent. The state is
associated with the presence of an AU-scale steady jet, and clear correlations
between the radio and X-ray intensities are observed \citep{cor13}. A major
question for the hard state is the geometry of the corona: While there is
significant evidence that the corona in the hard state is compact, it is quite
unclear whether, e.g., it is ADAF-like and quasi-spherical, hugs the disk, or
originates at the base of a jet \citep{cor00,shi11}.

\subsection{Controversy over the Location of the Disk's Inner
Edge}\label{sec:controversy}

In the thermal state, there is abundant evidence that the accretion disk is
truncated near the innermost stable circular orbit (ISCO) \citep[e.g.][]{gie04,
ste10,pen10,zhu12}. The standard paradigm for the faint hard state is that as
the luminosity decreases the inner edge of the disk recedes from the ISCO,
leaving a hot advection-dominated accretion flow (ADAF) or other coronal flow
\citep{nar94,nar08}. While there is good evidence that at very low luminosities
the disk is grossly truncated \cite[for a review, see][]{nar08}, the location
of the inner edge relative to the ISCO for luminosities in the range
$\sim0.1-10$\% of Eddington is a hotly-debated topic. With GX 339--4 as a
principal test bed, two methods have been widely used to estimate the radius
$R_{\rm in}$ of the inner edge of the disk in the hard state: (1) modeling the
component of emission reflected from the disk, principally the Fe K line; and
(2) fitting the continuum spectrum of the accretion disk. The former method,
which is addressed in the following section, is the central topic of this
paper.

Efforts to estimate the inner edge of the accretion disk in the low/hard state
via disk reflection go back farther, but the first strong indication that disks
may remain close to the ISCO in bright phases of the low/hard state was made by
\cite{mil06}.
Based on fits to the thermal component, a number of papers claim that
there is an optically thick disk that extends inward to the ISCO in the
hard state \citep{mil06,mil06c,ryk07,rei09b,rei10,reym10}.  This claim
is strongly contested by \cite{don07} and \cite{don10}; the claim is all the more 
questionable when one considers that self-consistent disk coronal models
\citep[e.g.,][]{ste09} return larger values of the inner-disk radius.
More recently, Miller and coworkers have invoked extreme values of the
spectral hardening factor in making the case for an untruncated
hard-state disk \citep{reym13,sal13}.  This evidence for the presence of
such a disk does not appear to us compelling given the difficulties of
obtaining accurate estimates of $R_{\rm in}$ by modeling a faint, cool
($kT \lesssim 0.2$~keV) thermal component that is strongly Comptonized
and cut off by interstellar absorption.

\subsection{Reflection Spectroscopy}\label{sec:refspec}

The reflection spectrum results from the reprocessing of high-energy
coronal photons in the optically-thick accretion disk. The result is a
rich spectrum of radiative recombination continua, absorption edges and
fluorescent lines, most notably the Fe K complex in the 6--8 keV energy
range. This reflected radiation leaves the disk carrying information on
the physical composition and condition of the matter in the strong
fields near the black hole. The Fe K emission line (and other
fluorescent lines) are broadened and shaped by Doppler effects, light
bending and gravitational redshift. By modeling the reflection spectrum,
one can estimate both the disk inclination and the dimensionless spin
parameter $a_* = cJ/GM^2$ ($-1 \le a_* \le 1$). In measuring $a_*$, one
estimates the radius of the inner edge of the accretion disk and
identifies it with the radius of the innermost stable circular orbit,
$R_{\rm ISCO}$, which simply and monotonically maps to $a_*$
\citep{bar72}. For the three canonical values of the spin parameter,
$a_* = +1$, 0 and -1, $R_{\rm ISCO} = 1M$, $6M$ and $9M$ (for $c = G =
1$), respectively.

The reflection model most widely used in the past for both general
application and measuring black hole spin is {\sc reflionx}
\citep{ros05}.  Recently, an improved reflection model has been
developed, {\sc relxill}
\footnote{http://www.sternwarte.uni-erlangen.de/research/relxill}, which
is based on the reflection code {\sc xillver} \citep{gar10,
  gar11,gar13a,gar14a}, and the relativistic line-emission code {\sc
  relline} \citep{dau10,dau13,dau14}.  Compared to {\sc reflionx}, {\sc
  relxill} incorporates a superior treatment of radiative transfer and
Compton redistribution, and it allows for the angular dependence of the
reflected spectrum.  Furthermore, by implementing the routines of the
photoionization code {\sc xstar} \citep{kal01}, {\sc relxill} provides
an improved calculation of the ionization balance.  At the same time,
limitations of the model include assuming that the density of the disk
is independent of vertical height, that the illuminating radiation
strikes the disk at a fixed angle of 45~deg, and that apart from Fe all
the elemental abundances are assumed to be solar.  The results presented
in this paper were derived using {\sc relxill} to model the
relativistically-blurred reflection component from the inner disk and
{\sc xillver} to model a distant reflector.

It is important to appreciate the faintness of the reflected features that are
crucial for probing effects in the regime of strong gravity, the features that
one relies on for estimating $R_{\rm in}$ and constraining black hole spin.
For example, in the spectrum of GX~339--4, even the most prominent feature, the
Fe K line, has a typical equivalent width of $\sim0.1$~keV, and the peak
intensity of the line is only about 10\% of the
local continuum (Section~\ref{sec:combined}). Sensitivity to such faint
features requires both high-count spectra and a well-calibrated detector.

\subsection{The Special Quality of This Study}\label{sec:special}

The principal detector aboard the {\it Rossi X-ray Timing Explorer}
(\rxte) was the Proportional Counter Array (PCA), which was comprised of
five nearly identical Proportional Counter Units (PCUs), each with an
effective area of 1600~cm$^2$ and with sensitivity from
2--60~keV. Despite the limited spectral resolution of the instrument
($\approx17$\% at 6~keV) the archive of PCA data amassed during the
\rxte\ mission (1995--2012) continues to be preeminent for the synoptic
study of stellar-mass black holes. A few-dozen bright black holes were
observed daily during their outburst cycles with typical exposure times
of a few ks. Some 15,000 individual spectra were obtained with a net
total exposure time of 30~Ms (1 year). In this paper, we report the
results of our analysis of six hard-state spectra of GX 339--4, each a
summation of dozens of individual exposures
(Section~\ref{sec:combined}). For the spectrum obtained at maximum
luminosity ($L/L_{\rm Edd}= 17$\%) with an exposure time of 46~ks, the
total number of counts is 40 million and the counts-per-keV in the
continuum at 6.4~keV is 4.4 million, while the total number of counts in
the Fe K line region (3--10~keV) is 28 million.

A limitation of the PCA, which has not allowed the implied statistical
precision to be realized in modeling data, has been the appreciable
$\sim1$\% uncertainties in the detector response \citep{jah06,
  sha12}. We have overcome this limitation by developing a calibration
tool, called \pcacorr, that increases the sensitivity of the \rxte\ PCA
detector to faint spectral features -- such as the Fe K line/edge -- by
up to an order of magnitude \citep{gar14b}. By applying \pcacorr\ to a
large number of spectra for three black holes, we found that the tool
improved the quality of all the fits, and that the improvement was
dramatic for spectra with $\gtrsim10^7$ counts. The tool allows one to
achieve a precision of $\sim0.1$\% rather than $\sim1$\%, thereby making
full use of spectra of bright sources with $\sim10^6$ counts per
channel.

Consequently, our study of the reflection spectrum of GX~339--4 greatly
improves on earlier studies using the PCA, such as that by \cite{pla15}.
A limitation of PCA data is its modest resolution, while its major
advantage is its freedom from the problematic effects of pileup, which
is commonly a serious problem in analyzing and interpreting data for
bright sources obtained using CCD detectors (see
Section~\ref{sec:spin}).  Another advantage of the PCA, which has only
recently been matched by \nustar, is its high-energy coverage, which
allows observations of both the Fe K region and the Compton hump using a
single detector.


%
%
\begin{figure*}
\centering
\includegraphics[scale=0.90,angle=0]{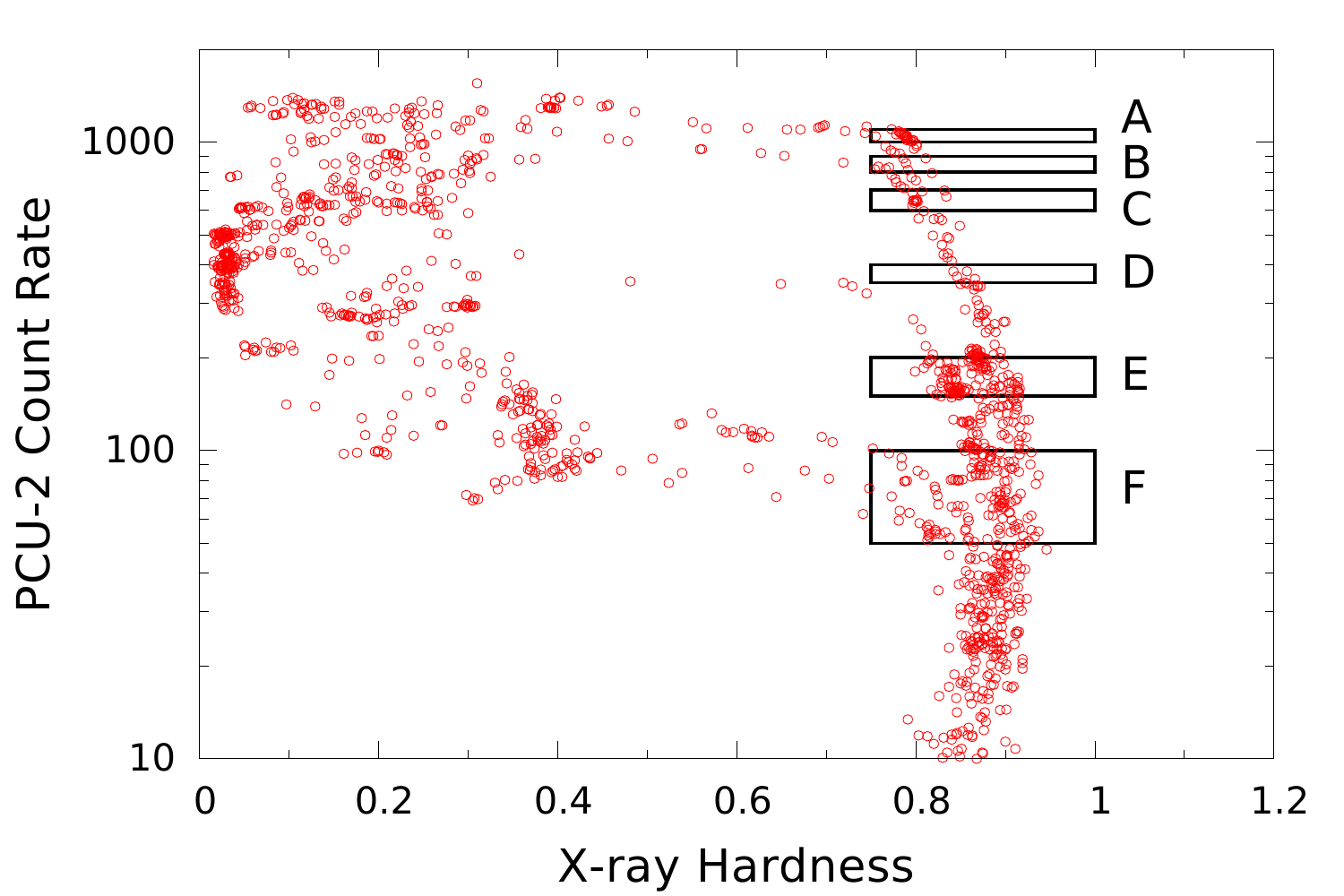}
\caption{ Hardness-intensity diagram for all PCU-2 \rxte\ observations
  of GX~339--4. The vertical axis shows the raw PCU-2 count rate (for
  reference, 1~Crab $\approx$ 2600 counts s$^{-1}$), a proxy for the
  X-ray intensity and luminosity. Plotted on the horizontal axis is the
  hardness ratio $HR$ defined as the ratio of source counts at
  8.6--18~keV to the counts at 5--8.6~keV. Following further the
  conventions of \cite{rem06}, the hard-state data considered
  exclusively in this paper are defined to have $HR>0.75$. The six boxes
  labeled A--F define the data sets we sum to create Spectra A--F, which
  are used in our analysis throughout the paper.  }
\label{fig:qdiagram}
\end{figure*}

This paper is organized as follows: Section~\ref{sec:obs} describes the
observations and data reduction, and Section~\ref{sec:combined} outlines our
procedure for combining the individual spectra into six composite spectra. The
luminosities of these spectra, which we refer to throughout as Spectra A--F,
range over an order of magnitude. Fitting the spectra individually, while
emphasizing the importance of correcting the data using the \pcacorr\ tool, is
the subject of Section~\ref{sec:fitcomb}. Our key results appear in
Section~\ref{sec:simultaneous}. Therein, we describe how we fit Spectra A--F
simultaneously, first fixing the spin parameter and letting the inner-disk
radius vary, and then allowing the spin parameter to vary while fixing the
inner radius at the ISCO. We discuss our results in
Section~\ref{sec:discussion} and offer our conclusions in
Section~\ref{sec:conc}.

\vspace{1.0cm}
\section{Observations and Data Reduction}\label{sec:obs}

Our reduction and analysis of the \rxte\ PCA data are detailed in
\cite{gar14b}, which follows the procedures of \cite{McClintock2006}. The data
were obtained in ``Standard 2'' mode and segmented into contiguous intervals
with exposure times ranging from 300~s to 5000~s. Background spectra, which
were derived using {\tt pcabackest} and the model {\tt
pca\_bkgd\_cmvle\_eMv20111129.mdl}, were subtracted from the data. Response
files were generated using {\tt pcarmf} (version 11.7) and the
energy-to-channel conversion table (version {\it e05v04}) described in
\cite{sha12}. Throughout, we analyze just the data collected using the
best-calibrated detector, PCU-2, which also provides the richest data set.

As a crucial final step, we apply the tool \pcacorr\ \citep{gar14b} to
the data and thereby calibrate the detector to a precision of $\sim
0.1\%$; we include a systematic error of this magnitude in all our data
analysis. This step greatly enhances the sensitivity of the detector
(Section~\ref{sec:pcacorr}) to the reflection features that are our
focus.

Figure~\ref{fig:qdiagram} presents a hardness-intensity diagram, PCA count rate
versus PCA hardness ratio $HR$, for all 1471 \rxte\ PCU-2 observations of
GX~339--4 obtained between 1996 July 26 and 2011 April 5. The hard-state data
considered exclusively in this paper are defined to have $HR>0.75$. In order
to boost the signal-to-noise, we define the six boxes A--F shown in
Figure~\ref{fig:qdiagram}. Each box contains a number of spectra, all of them
corresponding to roughly the same source intensity. We combine all the spectra
within a box using the procedures described in Section~\ref{sec:combined};
importantly, we do not combine spectra obtained during different outburst
cycles. Except for Box A, which is comprised of observations taken during
the 2002 outburst, all the other boxes contain observations taken during the
2010 outburst. Ultimately, we produce six master spectra (A--F), one for each 
box.


\section{Combining Spectra}\label{sec:combined}

We now outline our procedure for combining the individual spectra in a
box to create Spectra A--F in such a way as to eliminate small variations
in the power-law index under the assumption that the reflection features
are unaffected by small changes in the continuum. For each box
separately, we first fitted the individual spectra to a simple absorbed
power-law ({\tt Tbabs*powerlaw}) using
a fixed hydrogen column density of $N_\mathrm{H}=3\times 10^{21}$~cm$^{-2}$,
which is similar to the expected column in the direction of GX~339--4
\citep{kal05}. No evidence for a thermal component was found in any of the
spectra. The fits were performed in the 3--45~keV band where the Fe K features
are most pronounced\footnote{Inclusion or exclusion of the 4--7~keV band was
found to have a negligible effect on our final results.}. We then created
individual residual spectra (data counts minus model) and summed them, thereby
greatly enhancing the residual features present in these spectra.

\begin{figure*}
\centering
\includegraphics[scale=0.40,angle=0]{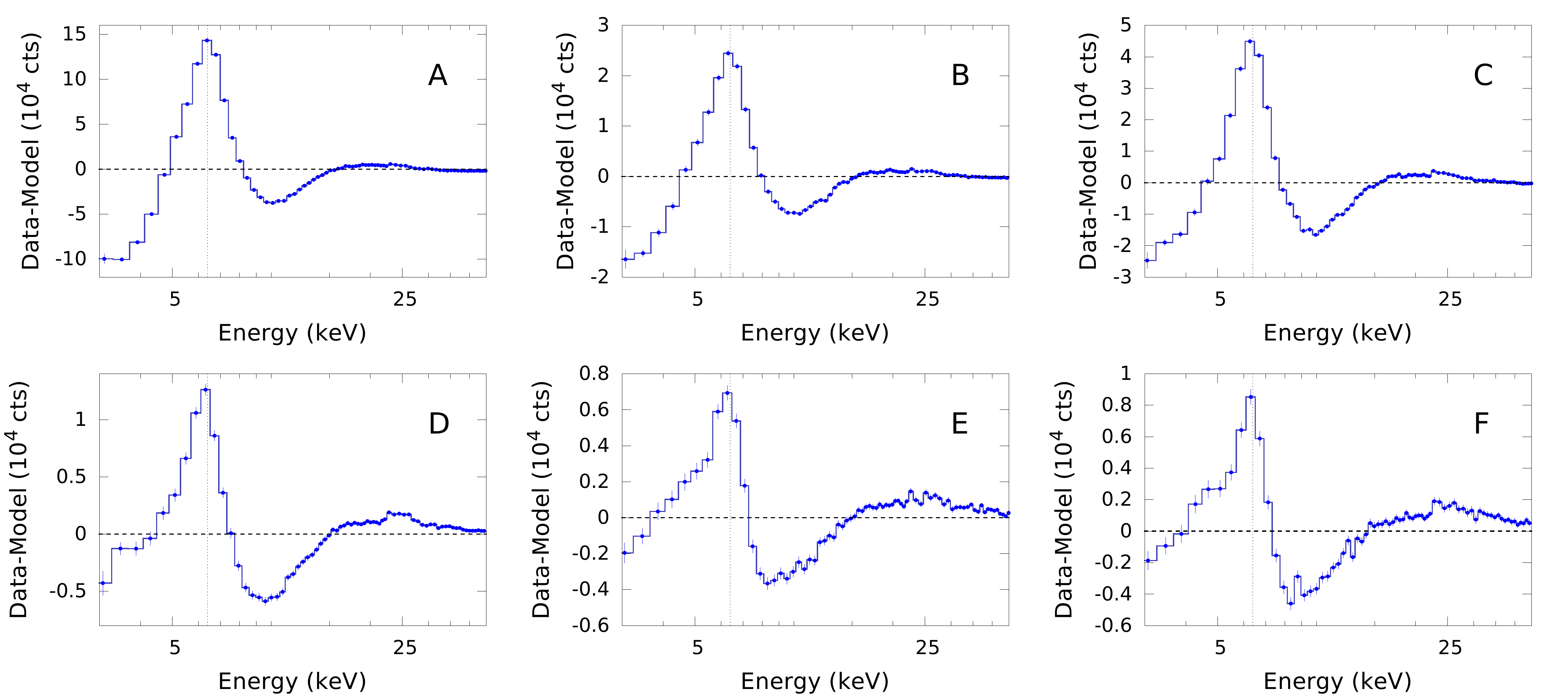}
\caption{
Residuals (data-minus-model) are computed for each box (defined in
Figure~\ref{fig:qdiagram}) by subtracting an absorbed power-law fit to
the individual spectra in the box. The extreme statistical precision
results from summing millions of counts, ranging from $\sim28$~million
for Box A to $\sim3$~million for Box F (3--10 keV). Note that the
spectra are all scaled differently; e.g., the peak signal for the
highest-luminosity Box A is $\sim 15$ times greater than for the
lowest-luminosity Box F. Remarkably, the appearance of the spectra is
quite insensitive to luminosity.
}
\label{fig:residuals}
\end{figure*}

Figure~\ref{fig:residuals} shows the residual spectra for the six
boxes. The striking features in each spectrum are the Fe K line and the
K edge, which are revealed with precision in these high signal-to-noise
spectra. Surprisingly, the overall structure of the residuals are in all
cases quite similar, despite the factor of $\sim 10$ spread in
luminosity (Figure~\ref{fig:qdiagram}). Upon closer examination,
however, one sees that the line width, the position of the edge, and the
shape of the Compton hump differ to some degree among the boxes. This
point is discussed in more detail in Section~\ref{sec:fit1}.


%
%

We now use these residuals to create Spectra A--F, which constitute our
prime data set. To be specific, consider the creation of Spectrum A: For
box A we generate a single template continuum spectrum using the average
values of the fit parameters $\Gamma$ and the normalization $N$.  This
spectrum is generated synthetically using the {\tt fakeit} task in {\sc
  xspec}; its net exposure time is the sum of the exposure times of all
the 23 individual spectra in box A. Finally, we add this continuum
spectrum to its corresponding summed residual spectrum to complete the
generation of Spectrum~A.

The resulting spectrum is superior to that obtained by simply summing
the individual spectra directly because it seamlessly eliminates the
effects of small differences in the power-law index and normalization
among the spectra.  The mean power-law parameters and other information
describing these six spectra, which are hereafter our focus, are
summarized in Table~\ref{tab:boxes}.  We compute the luminosity for a
spectrum using our model fluxes in the 1--100~keV band and assuming a
distance of $D=8$~kpc and black hole mass of $M=10~M_{\odot}$ (for
details, see Footnote $a$ to Table 1).

%
\begin{table*}
\caption{Properties of Spectra A--F and the boxes defined in
Figure~\ref{fig:qdiagram}}
\begin{center}
\label{tab:boxes}
\footnotesize
\begin{tabular}{ccccccc}
\hline
Spectrum & $L/L_\mathrm{Edd}$ (\%)$^a$ & Count Rate & \# Spectra & $<\Gamma>$ & $<N>$ & Exp. (ks) \\
\hline
A & 17.3 & $1000 - 1100$ & 23 & $1.72 \pm 0.01$ & $2.12 \pm 0.11$ & 45.7 \\
B & 14.2 &  $800 - 900$ &  7 & $1.75 \pm 0.03$ & $1.81 \pm 0.10$ & 10.3 \\
C & 11.9 &  $600 - 700$ & 11 & $1.69 \pm 0.01$ & $1.22 \pm 0.06$ & 27.0 \\
D &  7.9 &  $350 - 400$ &  7 & $1.61 \pm 0.60$ & $0.60 \pm 0.03$ & 15.7 \\
E &  3.9 &  $150 - 200$ & 18 & $1.52 \pm 0.01$ & $0.23 \pm 0.02$ & 24.6 \\
F &  1.6 &   $50 - 100$ & 43 & $1.59 \pm 0.08$ & $0.11 \pm 0.03$ & 72.7 \\
\hline
\end{tabular}
\end{center}
$^a$ Note. -- Eddington-scaled luminosities assuming $D=8$~kpc and 
$M=10~M_{\odot}$ (corresponding to $L_\mathrm{Edd}=1.25 \times
10^{39}~$~erg~s$^{-1}$), and based on the fluxes
computed over the 1--100~keV band using the model and fit parameters summarized
in Table~3.
\end{table*}
%

\section{Fitting spectra A--F individually}\label{sec:fitcomb}

The residual plots in Figure~\ref{fig:residuals} unambiguously demonstrate that
a strong reflected component is present, which is widely attributed to the
illumination of the disk by a hot corona. Invoking this paradigm, we proceed to
fit each of the six spectra using our physically-motivated reflection code
\relxill\ v0.2g \citep{gar14a}. As before, Galactic absorption is modeled using {\tt
Tbabs} with fixed column density ($N_\mathrm{H}=3\times 10^{21}$~cm$^{-2}$).
For the {\tt Tbabs} model \citep{wil00}, we used the \cite{and89} set of solar
abundances and the \cite{ver96} photoelectric cross sections.

\begin{figure*}
\centering
\includegraphics[scale=0.6]{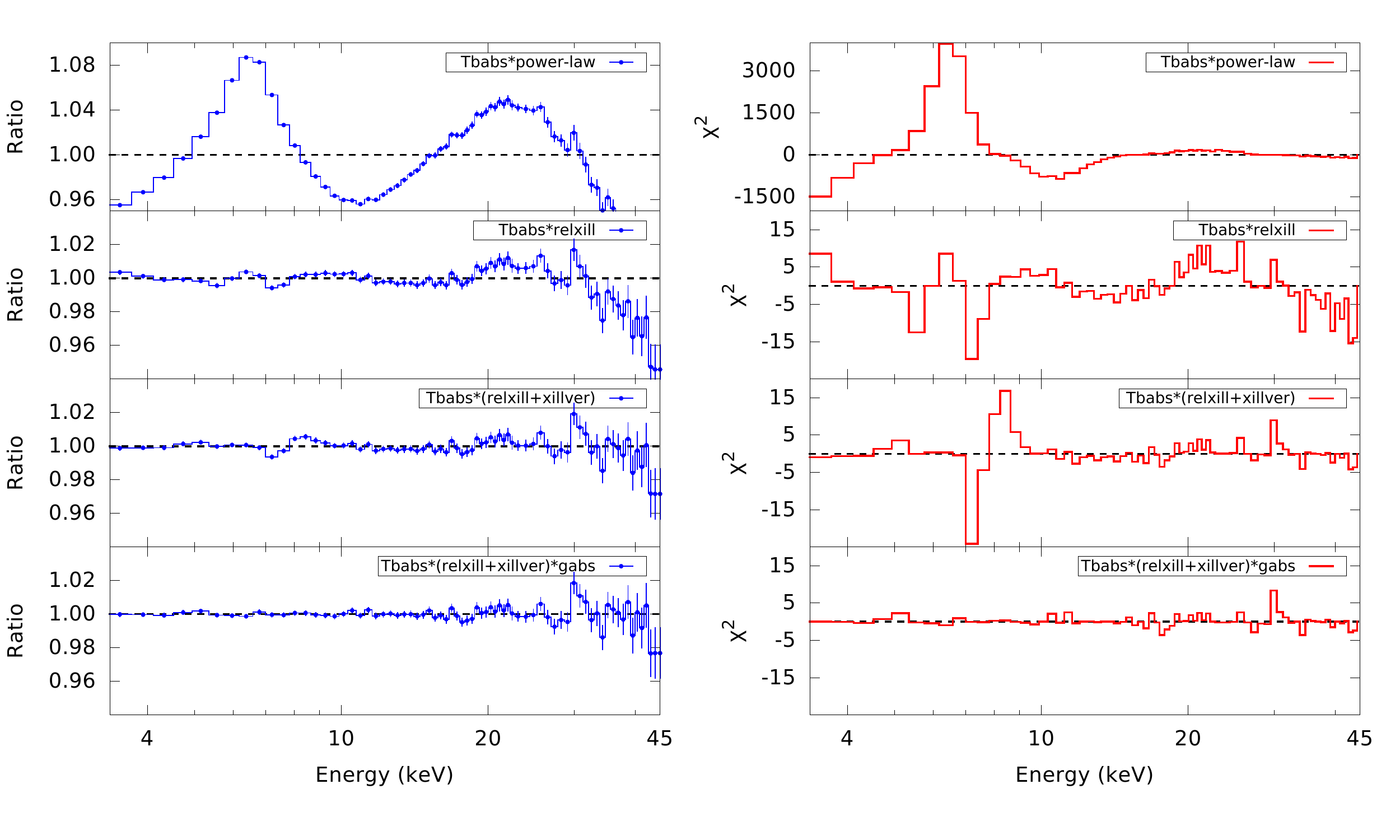}
\caption{ Data-to-model ratio ({\it left}) and contributions to $\chi^2$
  ({\it right}) for Spectrum~A resulting from fitting a sequence of four
  models.  From top to bottom, the models increase in completeness and
  competence, starting with an absorbed power-law model to which,
  incrementally, is added a blurred reflection component (\relxill);
  unblurred reflection (\xillver); and a Gaussian absorption line {\tt
    gabs}.  The comparable residual plots for the other five spectra are
  qualitatively similar.  }
\label{fig:progression}
\end{figure*}

We fit Spectra A--F in turn to a succession of four models; the final
adopted model in each case yields a good fit with
$\chi^2/\nu\sim1$. Table~\ref{tab:ifits} provides detailed information
on the quality of the fit for each spectrum and each model. In
Figure~\ref{fig:progression}, we show for Spectrum A with $4 \times
10^7$~counts -- the most challenging case -- residual plots for the
progression of the four models, which we now describe.


%
%

{\bf Model~0:} {\tt Tbabs*powerlaw}. An absorbed power-law model, which
is clearly deficient, prominently displays the principal reflection
features, the Fe K line/edge and Compton hump, in the residuals
(Figure~\ref{fig:progression}).

{\bf Model~1:} {\tt Tbabs*relxill}. A greatly improved fit to all six spectra
is achieved by replacing the power-law with our fully relativistic reflection
model. For simplicity and to achieve definiteness, we fix the spin to its
extreme value of $a_*=0.998$ and assume the canonical dependence of disk
emissivity with radius, namely $\propto r^{-3}$. This model already delivers
fits of reasonable quality (Table~\ref{tab:ifits}). Some pronounced
residual features remain, which are most evident for the most luminous case, Spectrum A
(Figure~\ref{fig:progression}). Specifically, two apparent absorption features
flank the Fe K line at $\sim 5.6$ and $\sim 7.2$~keV. Other features are also
present at higher energies in the region of the Compton hump ($\sim
20-45$~keV).

{\bf Model~2:} {\tt Tbabs*(relxill+xillver)}. The residuals are
significantly reduced by including an {\it unblurred} reflection
component via \xillver.  Physically, this reflector could be cold
material in a wind or in the outer region of a flared disk (see
below). The \xillver\ parameters are linked to those of \relxill\ with
two exceptions: The ionization parameter was fixed at its minimum value,
log~$\xi=0$, and the Fe abundance was fixed to solar (i.e.,
$A_\mathrm{Fe}=1$). Linking the Fe abundance results in a significantly
worse fit and a compromise value of abundance that is midway between the
low value required by the unblurred component and the super-solar value
required by the blurred component (see Section~\ref{sec:afe}). This
result, and the uncertain origin of the unblurred component, motivate
our choice of solar Fe abundance for the distant reflector.  We have no
good explanation for the different Fe abundances required in fitting the
blurred and unblurred reflection components.  Further discussion on the
Fe abundance is presented in Section~\ref{sec:afe}.

Inclusion of the \xillver\ component, which introduces only one new free
parameter, namely its normalization, quite significantly improves the fit to
all the spectra except Spectrum F, which has the fewest counts. While the
\xillver\ component improves the fit at low energies and in the region of the
Compton hump, a strong residual feature remains at $\sim 7.2$~keV
(Figure~\ref{fig:progression}).

{\bf Model~3:} {\tt Tbabs*(relxill+xillver)*gabs}. We model the
remaining residual feature near 7.2~keV phenomenologically as absorption
using a single Gaussian. The addition of this component improves the
fits substantially for Spectra A--C, i.e., those with many counts, while
it has only a marginal effect for Spectrum D and a negligible effect for
Spectra E and F (Table~\ref{tab:ifits}).  As expected, its importance is
greatest for Spectrum A where it completely eliminates the strong
7.2~keV residual feature (Figure~\ref{fig:progression}) and produces a
very good fit to this spectrum, despite its extreme statistical
precision ($4 \times 10^7$~total counts), with an allowance for
systematic error of only 0.1\% (Section~\ref{sec:obs}).  

It is important to note that the inclusion of the 7.2~keV feature has a
significant effect on some important model parameters.  In particular,
we find that including the Gaussian component (Model~3) increases the
inclination and decreases the inner-disk radius by about 4 degrees and
20\%, respectively, compared to excluding the component. The changes in
the other fit parameters are relatively much smaller. We adopt Model~3
as our fiducial model for all six spectra, thereby assuming that the
7.2~keV absorption feature has a physical origin. While it is beyond the
scope of this work to establish a definite physical interpretation of
the feature, we now briefly consider some plausible explanations.


%
\begin{table}
\caption{Statistics of the individual fits to Spectra A--F}
\begin{center}
\label{tab:ifits}
\footnotesize
\begin{tabular}{ccrcrr}
\hline
Spectrum & Model & $\chi^2$ & $\nu$ & $\chi^2_{\nu}$ & $\Delta\chi^2/\Delta\nu$ \\
\hline
A & 0 & 25094.90 & 69 & 363.694 &         \\
  & 1 &   299.07 & 63 &   4.747 & 4132.64 \\
  & 2 &   151.07 & 62 &   2.437 &  148.00 \\
  & 3 &    67.82 & 60 &   1.130 &   41.63 \\
\hline
B & 0 &  7653.27 & 69 & 110.917 &         \\
  & 1 &   117.00 & 63 &   1.857 & 1256.05 \\
  & 2 &    81.16 & 62 &   1.309 &   35.84 \\
  & 3 &    54.14 & 60 &   0.902 &   13.51 \\
\hline
C & 0 & 11849.29 & 70 & 169.276 &         \\
  & 1 &   142.44 & 64 &   2.226 & 1951.14 \\
  & 2 &    93.21 & 63 &   1.480 &   49.23 \\
  & 3 &    68.68 & 61 &   1.126 &   12.27 \\
\hline
D & 0 &  4880.44 & 70 &  69.721 &         \\
  & 1 &    89.91 & 64 &   1.405 &  798.42 \\
  & 2 &    41.36 & 63 &   0.657 &   48.55 \\
  & 3 &    35.69 & 61 &   0.585 &    2.84 \\
\hline
E & 0 &  2552.13 & 70 &  36.459 &         \\
  & 1 &   116.40 & 65 &   1.791 &  487.15 \\
  & 2 &    65.97 & 64 &   1.021 &   50.43 \\
\hline
F & 0 &  2311.81 & 70 &  33.026 &         \\
  & 1 &    63.17 & 65 &  0.9719 &  449.73 \\
  & 2 &    63.17 & 64 &  0.9870 &     0.0 \\
\hline
\end{tabular}
\end{center}
\end{table}
%

%
\subsection{The 7.2 keV absorption feature and the efficacy of the
  \xillver\ component}\label{sec:gabs}

We first consider the likely possibility that the absorption feature is largely
an artifact related to the uncertain energy resolution of the PCA. We then
discuss the one plausible physical explanation for the feature known to us,
namely that it is produced by absorption in a highly-ionized wind. Finally, we
consider the role of the \xillver\ component not only in modeling the residual
features near the Fe K line, but also its role in improving the fit quite
generally.

\subsubsection{On the accuracy of the PCA energy resolution}

\begin{figure}
\centering
\includegraphics[scale=0.55]{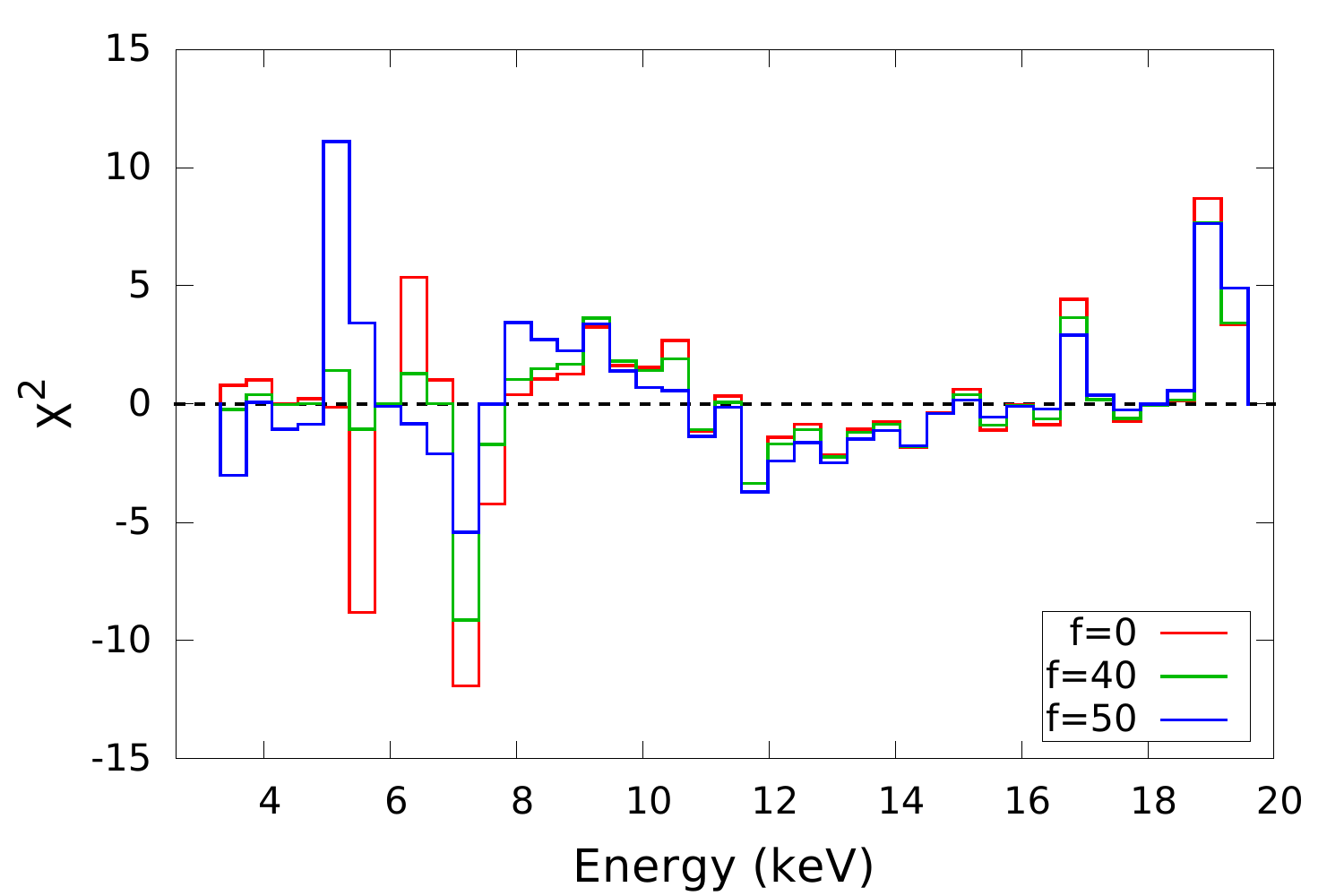}
\caption{Residual plots (data-model) illustrating the extreme
  sensitivity in fitting the Fe K line to the value assumed for the
  energy resolution of PCU-2.  The residuals are for fits to Spectrum A
  using Model~1 ({\tt Tbabs*relxill}). The resolution of the data have
  been slightly degraded by convolving them with a Gaussian whose
  smoothing width is characterized by the parameter $f$: The cases
  $f=40$ (green) and $f=50$ (blue) correspond to decreases in the
  resolution of 0.9\% and 1.5\%, respectively, and $f=0$ (red) is the
  unsmoothed case.}
\label{fig:smooth}
\end{figure}

The presence of residual absorption features bracketing the Fe line at $\sim
5.6$~keV and $\sim 7.2$~keV suggests the possibility that the PCA resolution
may be better than assumed in generating the PCU-2 response. We have explored
this possibility for Spectrum A. We test the effects of slight changes in the
value assumed for the detector resolution by moderately smoothing the data,
with the results shown in Figure~\ref{fig:smooth}. The smoothing is
accomplished using a Gaussian kernel operating over the detector channels; the
parameter $f$ specifies the width of the Gaussian as a percent of the channel
width. Accordingly, the curve in Figure~\ref{fig:smooth} labeled $f=0$ is
unsmoothed, while the curves labeled $f=40$ and $f=50$ correspond to degrading
the resolution of the data by 0.9\% and 1.5\% at 6.4~keV.

Although this approximate approach to artificially tuning the detector
resolution does not eliminate the residuals flanking the Fe line, it does
significantly reduce their strength. The test demonstrates that at this extreme
level of statistical precision the fit to a line feature is very sensitive to
the value assumed for the detector resolution. Specifically, if one assumes
that the nominal value of resolution for the unsmoothed case ($f=0$) at 6~keV
is 17.0\%, then the net resolution for $f=50$ is 17.3\% (i.e., the additional
blurring has a width of one-half channel, equivalent to $\sim 0.2$~keV at 6.4
keV, which is combined in quadrature with the nominal resolution width).
Meantime, the resolution of the PCA is not known to sufficient accuracy to
discriminate such fine differences (N.\ Shaposhnikov, private communication).
This suggests that the residuals near the Fe line may result from the PCA
resolution being slightly better than assumed in modeling the detector
response. However, this test is inconclusive. To properly assess the importance
of tuning the resolution, one must carry out a systematic analysis using the
PCA calibration software, which is beyond the scope of this paper.

\subsubsection{On the possibility that the feature originates in a highly
ionized wind}

If the 7.2~keV feature is not an instrumental artifact, a potentially
plausible explanation is that it originates in a highly ionized wind
that envelops the primary source. Disk winds have been observed in many
black hole binary systems, particularly at high accretion rates
\citep[e.g.,][]{pon12b, nei12, mil06b}. We investigated this possibility
by replacing the {\tt gabs} component by the photoionized warm absorber
model ({\tt warmabs}). We forced the Fe Ly$\alpha$ line at $\sim
6.9$~keV to be the dominant feature by setting the ionization parameter
to its maximum value (log~$\xi=5$). We linked the Fe abundances of {\tt
 warmabs} and \relxill\, while the abundances of all the other elements
remain at solar. The fitted blueshift of the Fe Ly$\alpha$ required to
model the 7.2~keV feature is $z=0.0576\pm 0.0101$, which corresponds to
an outflow velocity of $v=1.7\times10^4$~km~s$^{-1}$. The model provides
a good fit ($\chi^2_{\nu}=1.17$), which is very comparable to that
achieved using Model~3 (see the top-left panel of Figure~\ref{fig:tests}
for details and a comparison of the residuals). However, this
interpretation seems unlikely on physical grounds due to the extreme
column density required by the warm absorber, namely, $N_\mathrm{H}^{\rm
  abs}=(7.7\pm0.2)\times 10^{23}$~cm$^{-2}$.  If one links the {\tt
  warmabs} Fe abundance to that of the \xillver\ component (i.e.,
$A_\mathrm{Fe}=1$), the fit pegs at the hard limit of the {\tt warmabs}
model ($10^{24}$~cm$^{-2}$).

\subsubsection{On the inclusion of the unblurred reflection component}

Our initial motivation for including the unblurred \xillver\ component of
reflection was the presence of the $\sim 5.6$~keV and $\sim 7.2$~keV residual
features flanking the Fe line. However, as Figure~\ref{fig:progression} makes
clear, while the \xillver\ component effectively eliminated the low-energy
feature, it actually enhanced the 7.2~keV feature. The simplest ad hoc
phenomenological approach to eliminating both features is to include a pair of
Gaussian absorption lines in our model, which we did, fixing the widths of the
Gaussians to 0.01~keV and allowing the energy and strength of each line to vary.
While this model ({\tt Tbabs*relxill*gabs*gabs}) does clean up the $\sim
5.6$~keV and $\sim 7.2$~keV features, the quality of the fit,
$\chi^2_{\nu}=2.36$ ($\chi^2 = 139.25$ for 59 d.o.f.), is much poorer than
that achieved with Model~3, $\chi^2_{\nu}=1.13$ ($\chi^2 = 67.82$ for 60
d.o.f.), which uses \xillver\ and a single Gaussian. As the top-right panel of
Figure~\ref{fig:tests} makes clear for the stringent case of Spectrum A, the
latter model not only does a better job cleaning up the pair of targeted
residual features, it is also more effective at improving the fit at most other
energies as well. We consider this strong evidence for the presence in
GX~339--4 of a distant reflector.

\subsection{Comparing combined versus summed spectra}

As discussed in Section~\ref{sec:combined}, we combined the individual
spectra in order to increase our sensitivity to the reflection features
while minimizing the effects of jitter in the power-law index. To verify
the procedures we used in combining spectra, we compare the results of
fitting our Spectrum A to those obtained by fitting a spectrum created
by summing directly all the spectra in Box A\footnote{The effects of
  changes in the detector response are negligible because the 23
  observations were all made during a 10-day interval.}. Applying
\pcacorr\ and fitting both spectra with our adopted model (i.e., {\tt
  Tbabs*(relxill+xillver)*gabs}), we find that the model parameters are
all consistent. However, the fit to the summed spectrum is of
significantly lower quality ($\Delta\chi^2 = 62.97$) than the fit to
Spectrum A created using the procedures described in
Section~\ref{sec:combined}. Furthermore, as shown in the bottom-left
panel of Figure~\ref{fig:tests}, the residuals for the summed spectrum
are larger in almost every energy channel. These results demonstrate
that our method of combining the individual spectra significantly
improves the quality of the fit.

\subsection{The importance of applying the \pcacorr\ tool}\label{sec:pcacorr}

\begin{figure*}
\centering
\includegraphics[scale=0.55]{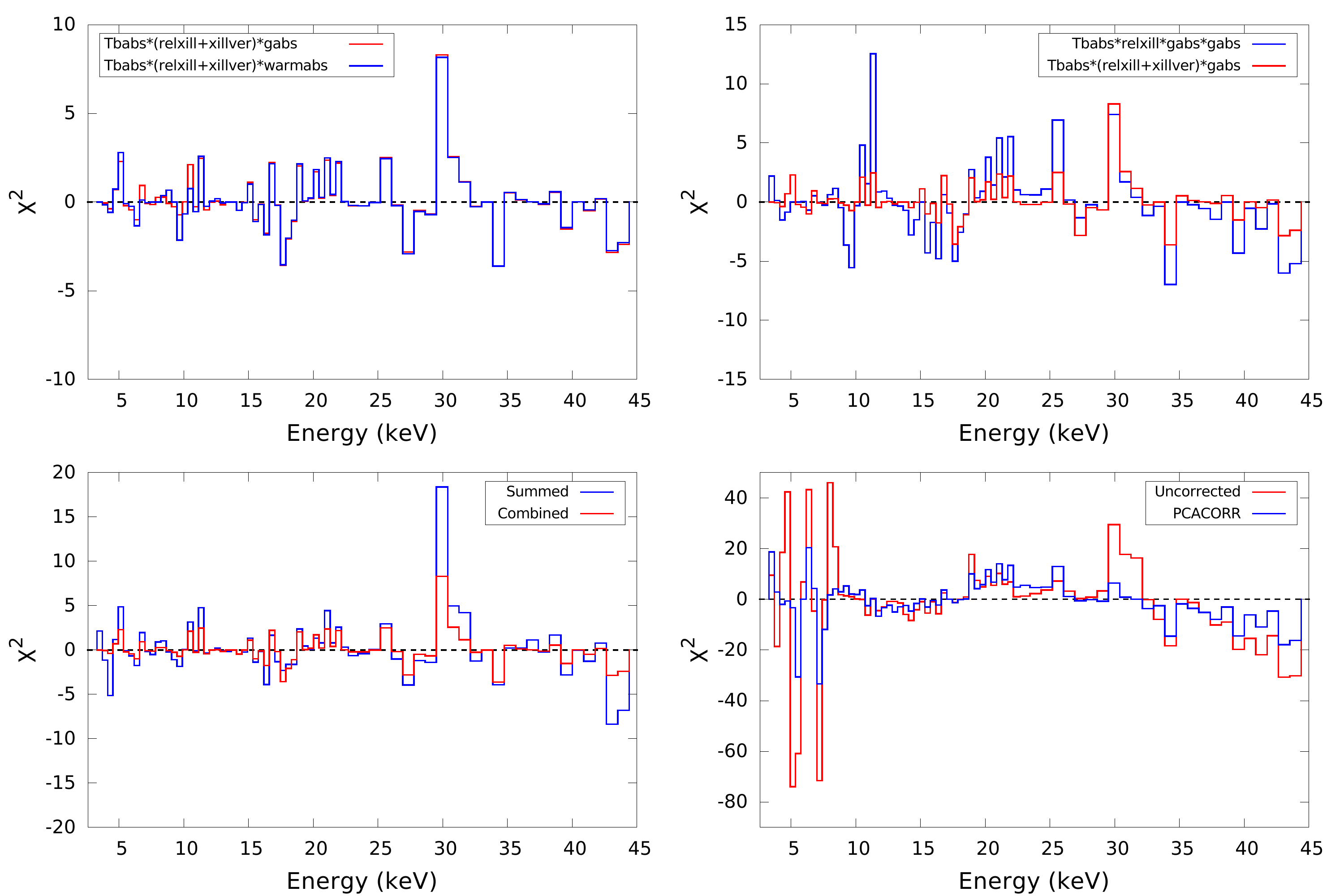}
\caption{
Comparative plot showing contributions to the total $\chi^2$ (data-model) for
each channel for fits to Spectrum~A:
({\it top-left}) The red curve was computed for our adopted Model~3 ({\tt
  Tbabs*(relxill+xillver)*gabs})  and is
identical to the plot shown in the lower-right panel in
Figure~\ref{fig:progression}. The blue curve is for an alternative model that
substitutes the warm absorber model {\tt wabs} for {\tt gabs} in our adopted
Model~3. The alternative model provides a good fit to the data: $\chi^2 =
69.06$ for 59 d.o.f.\ ($\chi^2_{\nu}=1.130$); compare $\chi^2$ values for
Model~3 in Tables~\ref{tab:fit1} and \ref{tab:fit2}.
({\it top-right}) The red curve was computed for our adopted Model~3 and is
identical to the plot shown in the lower-right panel in
Figure~\ref{fig:progression}. The blue curve is for an alternative model that
replaces the unblurred (\xillver) reflection component by a second Gaussian
absorption line at $\sim 5.6$~keV (blue), which results in a distinctly
inferior fit. As this residual plot shows, Model~3 performs better at almost
all energies.
({\it bottom-left}) Fits of our adopted Model~3 for two cases: (1) A direct sum
of the 23 spectra in Box A (blue), and (2) a fit to Spectrum A (red), which was
prepared by combining the spectra according to the procedures described in
Section~\ref{sec:combined}. While the model parameters are consistent for the
two cases, Spectrum A provides a superior fit, as this comparison of the
residuals makes clear.
({\it bottom-right}) Fits using Model~1 ({\tt Tbabs*relxill}) for two cases:
(1) The 23 individual spectra that comprise Spectrum A are corrected using the
\pcacorr\ tool (blue) and (2) they are left uncorrected (red). In this instance
only, all systematic errors have been zeroed to most clearly illustrate the
effect of applying \pcacorr\ to the data.
}
\label{fig:tests}
\end{figure*}

As fully described in \cite{gar14b} and discussed in
Section~\ref{sec:special}, the \pcacorr\ tool greatly reduces the
effects of instrumental features in PCA spectra, thereby making it
possible to achieve good fits to high-count spectra at the 0.1\% level
of statistical precision. The bottom-right panel in
Figure~\ref{fig:tests} demonstrates the importance of applying the tool
to Spectrum A, with its $4 \times 10^7$ counts. The figure compares
residuals for a fit to uncorrected data to one using \pcacorr-corrected
data. To most clearly illustrate the power of \pcacorr, we use Model~1
({\tt Tbabs*relxill}) and set the systematic errors to zero. Concerning
the $\sim 5.6$~keV and $\sim 7.2$~keV features flanking the Fe K line,
we note that they are present prior to the application of \pcacorr,
which confirms that they are not introduced by the correction.  The key
message of the bottom-right panel in Figure~\ref{fig:tests} is the
degree to which \pcacorr\ diminishes these features and others,
especially those below 10~keV and the one near 30 keV, a feature that is
likely related to the detector Xe K-edge \citep{sha12,gar14b}.


%
%

%
%


\begin{table*}
\caption{Results for {\sc jf-i}: Fit parameters for Model~3,
{\tt const*Tbabs*(relxill+xillver)*gabs}, with fixed maximum
spin $a_*$ and free $R_{\rm in}$.}
\begin{center}
\label{tab:fit1}
\footnotesize
\begin{tabular}{lccccccc}
\hline
Model & Parameter & Spectrum~A & Spectrum~B & Spectrum~C & Spectrum~D & Spectrum~E & Spectrum~F \\
\hline
{\tt Tbabs}   & $N_\mathrm{H}$ (cm$^{-2}$)& \multicolumn{6}{c}{$(7.0 \pm 1.0)\times 10^{21}$} \\
{\tt relxill} & $a_*$  & \multicolumn{6}{c}{$0.998$} \\
{\tt relxill} & $i$ (deg)         & \multicolumn{6}{c}{$                                 48.4 \pm 1.1 $} \\
{\tt relxill} & $A_\mathrm{Fe}$   & \multicolumn{6}{c}{$                          4.6^{+ 0.5}_{- 0.3} $} \\
{\tt relxill} & $N_\mathrm{r}$    & \multicolumn{6}{c}{$                       1.48^{+ 0.05}_{- 0.03} $} \\
{\tt gabs}    & $E$ (keV)         & \multicolumn{6}{c}{$                       7.19^{+ 0.04}_{- 0.06} $} \\
\hline
{\tt Constant}& & 1 & $0.91 \pm 0.04 $ & $0.71 \pm 0.02 $ & $0.36 \pm 0.01 $ & $0.16\pm 0.01 $ & $0.08 \pm 0.01 $ \\
{\tt relxill} & $R_\mathrm{in}$ ($R_\mathrm{ISCO}$) & $1.7^{+ 0.2}_{- 0.3} $ & $1.5^{+ 0.4}_{- 0.2} $ & $1.8^{+ 0.1}_{- 0.2} $ & $2.1^{+ 0.3}_{- 0.5} $ & $2.7^{+ 0.5}_{- 1.5} $ & $3.7^{+ 0.7}_{- 1.0} $ \\
{\tt relxill} & $\Gamma$ &         $1.620 \pm 0.013 $ & $1.682 \pm 0.016 $ & $1.672 \pm 0.013 $ & $1.628 \pm 0.015 $ & $1.588 \pm 0.010 $ & $1.648^{+ 0.007}_{- 0.012} $ \\
{\tt relxill} & log$\xi$ &        $3.31^{+ 0.03}_{- 0.07} $ & $3.24 \pm 0.07 $ & $3.12^{+ 0.07}_{- 0.03} $ & $3.031^{+ 0.020}_{- 0.013} $ & $2.02 \pm 0.17 $ & $2.05^{+ 0.26}_{- 0.10} $ \\
{\tt relxill} & $E_\mathrm{cut}$ & $97^{+ 3}_{- 5} $ & $129 \pm 10$ & $179 \pm 14 $ & $660^{+ 130}_{- 170} $ & $> 840$ & $> 890$ \\
{\tt relxill} & $R_\mathrm{f}$            & $0.21 \pm 0.02 $ & $0.22 \pm 0.03 $ & $0.21 \pm 0.03 $ & $0.34^{+ 0.04}_{- 0.05} $ & $0.31 \pm 0.03 $ & $0.31^{+ 0.02}_{- 0.03} $ \\
{\tt xillver} & $N_\mathrm{x}$   & $0.27^{+ 0.02}_{- 0.03} $ & $0.25 \pm 0.04 $ & $0.25 \pm 0.04 $ & $0.34^{+ 0.04}_{- 0.05} $ & $< 0.05$ & $ < 0.03$ \\
{\tt gabs}    & Strength         & $0.021^{+ 0.009}_{- 0.007} $ & $0.029^{+ 0.017}_{- 0.012} $ & $0.036 \pm 0.016 $ & $ 0.05^{+ 0.07}_{- 0.03} $ & $0.08^{+ 0.16}_{- 0.05} $ & $0.14^{+ 0.10}_{- 0.05} $ \\
\hline
   &  $L/L_\mathrm{Edd} (\%)$ & 17.3 & 14.2 & 11.9 & 7.9 & 3.9 & 1.6 \\
\hline
 & $\chi^2$       & \multicolumn{6}{c}{402.49} \\
 & $\nu$          & \multicolumn{6}{c}{379} \\
 & $\chi^2_{\nu}$ & \multicolumn{6}{c}{1.06} \\
\hline
\end{tabular}
\end{center}
\tablecomments{For the given model components, the parameters from top to bottom are:
  hydrogen column density ($N_\mathrm{H}$); dimensionless spin parameter ($a_* = cJ/GM^2$,
  where $J$ is the angular momentum of the black hole); inclination of the inner disk ($i$);
  iron abundance with respect to its solar value ($A_\mathrm{Fe}$); normalization of the
  blurred reflection component plus power-law continuum ($N_\mathrm{r}$); energy of the
  absorption Gaussian centroid ($E$); constant multiplicative factor between spectra;
  inner-disk radius ($R_\mathrm{in}$), with $R_\mathrm{ISCO}=1.237 R_g$
  for $a_*=0.998$ ($R_g = GM/c^2$);
  power-law photon index ($\Gamma$); log of the ionization
  parameter ($\xi = 4\pi F_\mathrm{x}/n$, where $F_\mathrm{x}$ is the ionizing flux and
  $n$ is the gas density); high-energy cutoff ($E_\mathrm{cut}$); reflection fraction
  ($R_\mathrm{f}$, ratio of the reflected flux to that in the power-law, in the 20--40~keV band);
  normalization of the distant (unblurred) reflection ($N_\mathrm{x}$); strength of the
  absorption Gaussian; X-ray luminosity in terms of Eddington (see notes in Table~\ref{tab:boxes});
  goodness of the fit ($\chi^2$); number of degrees of freedom ($\nu$); goodness of the
  fit per degree of freedom ($\chi^2_{\nu} = \chi^2/\nu$). Uncertainties are based on a 90\%
  confidence level.}
\end{table*}
%

%
\section{Fitting Spectra A--F simultaneously}\label{sec:simultaneous}

In order to achieve the strongest possible constraints on the key model
parameters, we fitted Spectra A--F simultaneously. This composite data set is
an assemblage of 106 individual PCA/PCU-2 spectra of GX~339--4 in the hard
state. The total number of counts is 77~million, 34~million of which are in the
3--10~keV Fe K band.

All fits are performed using Model~3 ({\tt
const*Tbabs*(relxill+xillver)*gabs}). To accommodate the order-of-magnitude
range of luminosity, we included a normalization constant that is unity for
Spectrum A and floats for Spectra B--F. Where sensible, key physical parameters
are tied: the spin $a_*$ of the black hole; inclination $i$ of the system;
line-of-sight column density $N_{\rm H}$; Fe abundance $A_\mathrm{Fe}$; and the
normalization of the relativistic reflection component $N_\mathrm{r}$ (see
Table~\ref{tab:fit1}). Given the uncertain origin of the absorption feature
near 7.2~keV, which we model as a Gaussian, we also tie the central energy of
this component while allowing its normalization to vary from spectrum to
spectrum. The width of the Gaussian was fixed at 0.01~keV in all the spectra.
Like the Gaussian normalization, all other model parameters are allowed to vary
independently.

Despite the extreme signal-to-noise of the composite spectrum, we must
impose some additional assumptions in order to simultaneously constrain
both the spin parameter $a_*$ and the radius of the inner edge of the
accretion disk $R_{\rm in}$. This is because these two quantities, which
manifest almost indistinguishably in the red wing of the Fe K line
profile, are extremely degenerate \citep{dau13}. Therefore, we have
conducted two complementary analyses, which we will refer to as {\sc
  jf-i} and {\sc jf-ii}, where {\sc jf} signifies that these are joint
fits (i.e., simultaneous) to Spectra A-F (rather than the fits to
Spectrum A alone that are featured in earlier sections). 
For {\sc jf-i}, our aim is to constrain $R_{\rm in}$, and we therefore
keep the spin fixed at its maximum value of $a=0.998$.
For {\sc jf-ii}, we tie the spin parameter for the six spectra and fit
for it, while fixing the inner edge of the disk at the ISCO; i.e.,
$R_{\rm in} = R_{\rm ISCO}$.  We follow the guidelines of \citet{fab12a}
and fix the emissivity index to its canonical value of 3; for both {\sc
  jf-i} and {\sc jf-ii}, we do not attempt to fit for this parameter
because of the PCA's limited spectral resolution.  This choice is
motivated by several {\sc jf-i} and {\sc jf-ii} tests we performed that
returned values of the emissivity index that were always $\lesssim 4$.

The number of free fitting parameters is large, 52 for {\sc jf-i} and 47 for
{\sc jf-ii}. The complexity of the analysis dictated our approach: We performed
Markov Chain Monte-Carlo (MCMC) runs using the {\sc emcee-hammer} Python
package \citep{dfm13}, which implements affine-invariant sampling. MCMC methods
are powerful for high-dimensional analysis. Specifically, they enable an
efficient exploration of parameter space and determine a posterior probability
structure for the model of interest.


\begin{table*}
\caption{Results for {\sc jf-ii}: Fit parameters for Model~3,
  {\tt const*Tbabs*(relxill+xillver)*gabs},
  with $a_*$ free and $R_{\rm in}=R_{\rm ISCO}$.}
\begin{center}
\label{tab:fit2}
\footnotesize
\begin{tabular}{lccccccc}
\hline
Model & Parameter & Spectrum~A & Spectrum~B & Spectrum~C & Spectrum~D & Spectrum~E & Spectrum~F \\
\hline
{\tt Tbabs}   & $N_\mathrm{H}$ (cm$^{-2}$)& \multicolumn{6}{c}{$(5.9^{+ 0.6}_{- 1.9})\times 10^{21}$} \\
{\tt relxill} & $a_*$  & \multicolumn{6}{c}{$                       0.95^{+ 0.03}_{- 0.05} $} \\
{\tt relxill} & $i$ (deg)         & \multicolumn{6}{c}{$                         47.8^{+ 0.9}_{- 1.4} $} \\
{\tt relxill} & $A_\mathrm{Fe}$   & \multicolumn{6}{c}{$                          5.4^{+ 1.9}_{- 0.5} $} \\
{\tt relxill} & $N_\mathrm{r}$    & \multicolumn{6}{c}{$                       1.44^{+ 0.04}_{- 0.08} $} \\
{\tt gabs}    & $E$ (keV)         & \multicolumn{6}{c}{$                                7.23 \pm 0.08 $} \\
\hline
{\tt Constant}& & 1 & $0.90^{+ 0.03}_{- 0.04} $ & $0.71^{+ 0.02}_{- 0.03} $ & $0.37 \pm 0.02 $ & $0.17\pm 0.01$ & $0.08 \pm 0.01 $ \\
{\tt relxill} & $R_\mathrm{in}$ ($R_\mathrm{ISCO}$) & \multicolumn{6}{c}{1} \\
{\tt relxill} & $\Gamma$ &         $1.604^{+ 0.010}_{- 0.027} $ & $1.658 \pm 0.018 $ & $1.651^{+ 0.015}_{- 0.022} $ & $1.62^{+ 0.02}_{- 0.04} $ & $1.578^{+ 0.009}_{- 0.013} $ & $1.637^{+ 0.009}_{- 0.013} $ \\
{\tt relxill} & log$\xi$ &        $3.33 \pm 0.03 $ & $3.35 \pm 0.04 $ & $3.16^{+ 0.10}_{- 0.05} $ & $3.05^{+ 0.04}_{- 0.02} $ & $1.96^{+ 0.12}_{- 0.21} $ & $2.0^{+ 0.2}_{- 0.2} $ \\
{\tt relxill} & $E_\mathrm{cut}$ & $92^{+ 2}_{- 6} $ & $118 \pm 8$ & $160^{+ 12}_{- 16} $ & $440^{+ 230}_{- 110} $ & $ >830 $ & $>940 $ \\
{\tt relxill} & $R_\mathrm{f}$            & $0.20 \pm 0.01 $ & $0.20^{+ 0.02}_{- 0.01} $ & $0.20 \pm 0.02 $ & $0.27^{+ 0.03}_{- 0.05} $ & $0.25^{+ 0.05}_{- 0.03} $ & $0.28^{+ 0.02}_{- 0.04} $ \\
{\tt xillver} & $N_\mathrm{x}$   & $0.23^{+ 0.05}_{- 0.02} $ & $0.26 \pm 0.04 $ & $0.24 \pm 0.05 $ & $0.13 \pm 0.06 $ & $0.08 \pm 0.06 $ & $< 0.09$ \\
{\tt gabs}    & Strength         & $0.024^{+ 0.009}_{- 0.007} $ & $0.028^{+ 0.018}_{- 0.011} $ & $0.037^{+ 0.016}_{- 0.012} $ & $0.04^{+ 0.03}_{- 0.02} $ & $0.025 \pm 0.014 $ & $ 0.020^{+ 0.017}_{- 0.009} $ \\
\hline
   &  $L/L_\mathrm{Edd} (\%)$ & 17.3 & 14.2 & 11.9 & 7.9 & 3.9 & 1.6 \\
\hline
 & $\chi^2$       & \multicolumn{6}{c}{418.66} \\
 & $\nu$          & \multicolumn{6}{c}{384} \\
 & $\chi^2_{\nu}$ & \multicolumn{6}{c}{1.09} \\
\hline
\end{tabular}
\end{center}
\tablecomments{For the given model components, the parameters from top to bottom are:
hydrogen column density ($N_\mathrm{H}$); dimensionless spin parameter ($a_* = cJ/GM^2$,
where $J$ is the angular momentum of the black hole); inclination of the inner disk ($i$);
iron abundance with respect to its solar value ($A_\mathrm{Fe}$); normalization of the
blurred reflection component plus power-law continuum ($N_\mathrm{r}$); energy of the
absorption Gaussian centroid ($E$); constant multiplicative factor between spectra;
inner-disk radius ($R_\mathrm{in}$); power-law photon index ($\Gamma$); log of the ionization
parameter ($\xi = 4\pi F_\mathrm{x}/n$, where $F_\mathrm{x}$ is the ionizing flux and
$n$ is the gas density); high-energy cutoff ($E_\mathrm{cut}$); reflection fraction
($R_\mathrm{f}$, ratio of the reflected flux to that in the power-law, in the 20--40~keV band);
normalization of the distant (unblurred) reflection ($N_\mathrm{x}$); strength of the
absorption Gaussian; X-ray luminosity in terms of Eddington (see notes in Table~\ref{tab:boxes});
goodness of the fit ($\chi^2$); number of degrees of freedom ($\nu$); goodness of the
fit per degree of freedom ($\chi^2_{\nu} = \chi^2/\nu$). Uncertainties are based on a 90\%
confidence level.}

\end{table*}
%

%
\begin{figure}
\centering
\includegraphics[scale=0.4,trim={0 2.8cm 0 0}]{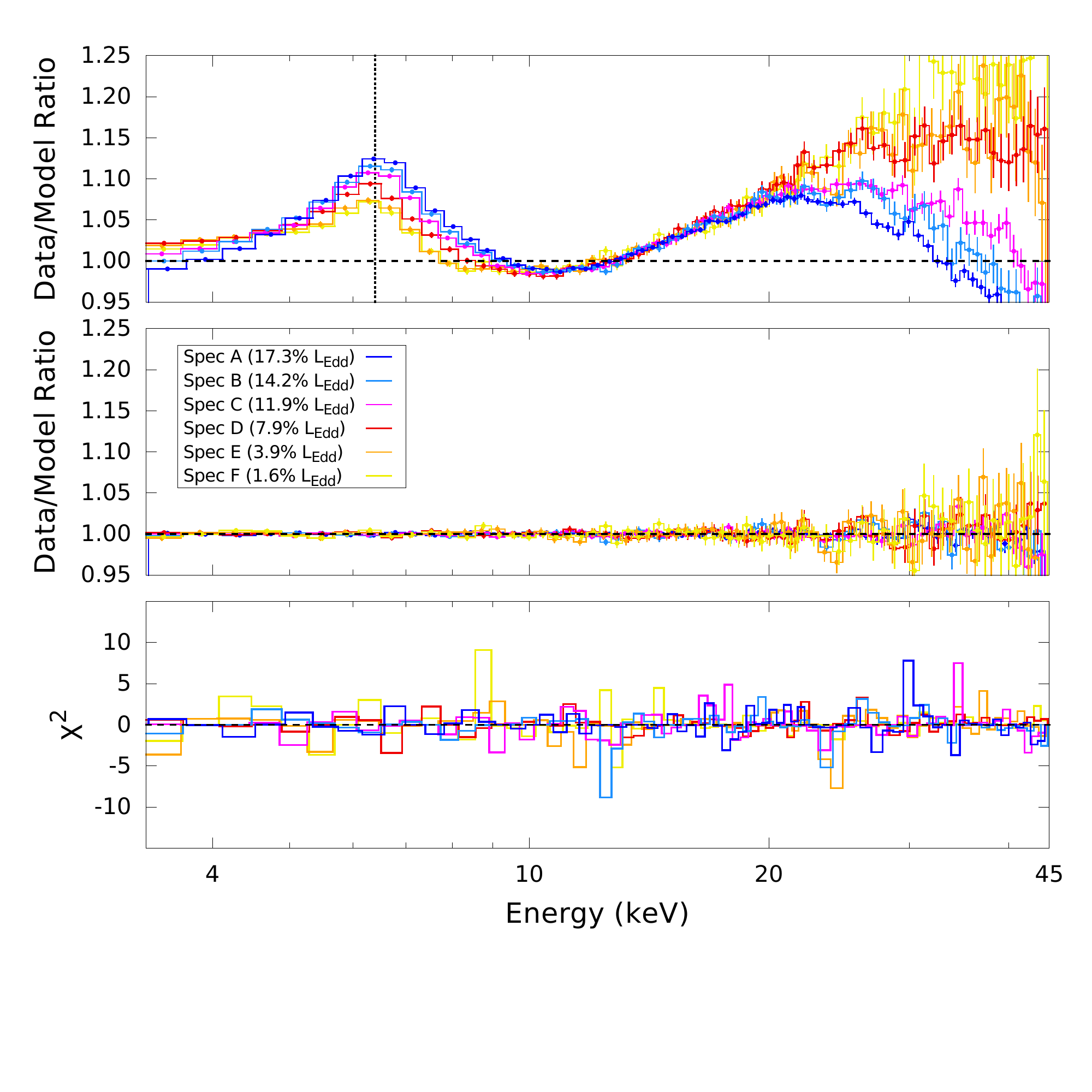}
\caption{ A simultaneous fit to Spectra A--F comprising a total of 106
  individual spectra and 77 million counts.  ({\it top}) Ratio plots for
  a joint fit ({\sc jf-i}) using Model~0, the simple absorbed power-law
  model, which strikingly reveal the principal signatures of
  reflection. ({\it middle}) Ratio plots for these same data obtained by
  fitting our canonical Model~3 ({\tt
    Tbabs*(relxill+xillver)*gabs}). This model produces an excellent fit
  with $\chi_{\nu}^2=1.06$.  ({\it bottom}) Contributions to the total
  $\chi^2$ (data-model), again for Model~3 and the same data.  }
\label{fig:fit1}
\end{figure}
%

\subsection{Joint Fit I: Fixed spin and variable inner radius}\label{sec:fit1}

A principal goal of our study is to track the radius of the inner edge
of the disk $R_{\rm in}$ as the luminosity varies by an order of
magnitude (i.e., over the range 1.6--17\% of Eddington;
Table~\ref{tab:boxes}). As discussed in Section~\ref{sec:controversy}, a
question of great interest is whether the inner disk is truncated in the
hard state at low luminosities and, if so, to what extent. In order to
be maximally sensitive to a disk that is only slightly truncated, we fix
the spin to its maximum allowed value, namely $a_*=0.998$. In so doing,
our focus is on determining how $R_{\rm in}$ trends with luminosity
rather than obtaining accurate estimates of this parameter. We note that
most spin determinations in the literature (which assume $R_{\rm in} =
R_{\rm ISCO}$) suggest that the spin is high (see
Section~\ref{sec:spin}).

Figure~\ref{fig:fit1} shows the fit residuals for {\sc jf-i} for two
cases: (1) The top panel shows a data-to-model ratio for a fit to
Model~0, i.e., the simple absorbed power-law model used to produce the
residual plot for a fit to Spectrum A only (which is shown in the
left-top panel of Figure~\ref{fig:progression}). This simple fit
prominently displays the reflection features, which are strong for all
six spectra. The profile of the Fe K line shows moderate variations
among the spectra. As the luminosity increases, so does the intensity of
the line (cf. Figure~\ref{fig:residuals}). Additionally, the blue wing
of the line extends to higher energies for the high luminosity spectra,
which could be evidence for a shift in the Fe K edge caused by an
increase in the ionization of the gas.  There are obvious changes at
high energies among the spectra, which are likely due to the evolution
of the high-energy cutoff with luminosity. (2) Of chief interest, the
middle and bottom panels of Figure~\ref{fig:fit1} show the residuals for
our adopted Model~3 (for $a_*=0.998$).

Model~3 performs remarkably well for all spectra as indicated by the
goodness of fit, $\chi^2_{\nu}=1.06$, and also by the uniformly ergodic
appearance of the residuals across the energy band. The fit results are
summarized in Table~\ref{tab:fit1}. A key result, which is discussed in
detail in the following section, is evidence that the disk is moderately
-- but significantly -- truncated at the lowest luminosities that can be
effectively explored using \rxte~(i.e., at a few percent of Eddington).

\vspace{0.5cm}
\subsection{Joint Fit II: Constraining the spin of the black hole}\label{sec:fit2}

In order to obtain constraints on black hole spin using either leading
method, reflection spectroscopy or continuum fitting, one must assume
that $R_{\rm in} = R_{\rm ISCO}$ \citep[e.g.,][]{rey14,mcc14}. In
performing {\sc jf-ii}, we make this assumption {\it for all six
  spectra} in order to constrain the spin of the black hole. Doing
so allows us to obtain a precise estimate of spin: $a_*=0.95^{+0.03}_{-0.05}$ at
90\% confidence. As the summary of results in Table~\ref{tab:fit2}
shows, the other parameters are quite close to those obtained in {\sc
  jf-i} (Table~\ref{tab:fit1}), and the goodness of fit is of very
comparable quality: $\chi^2_{\nu}=1.09$. Given the extreme
statistical precision, the residual spectra (data/model and $\chi^2$)
for {\sc jf-ii} (not shown) are essentially indistinguishable by eye
from the spectra for {\sc jf-i}, which are shown in
Figure~\ref{fig:fit1}.

Concerning our spin estimate and the fundamental assumption that $R_{\rm in} =
R_{\rm ISCO}$, we again note that {\sc jf-i} provides evidence for disk
truncation at our lowest luminosities\footnote{Furthermore, our {\sc jf-i}
results are consistent with some small degree of truncation even for our
higher-luminosity data.}.  Meantime, these low-luminosity spectra are included
in computing our single, tied {\sc jf-ii} estimate of spin. The incorporation
of disk truncation effects (which we have ignored) would imply an even higher
spin value than is quoted above.

\section{Discussion}\label{sec:discussion}

The MCMC runs utilized 120 ({\sc jf-i}) and 100 ({\sc jf-ii}) ``walkers,'' each
navigating a chain with a length of 100,000 elements, after having been
initialized in a cluster distributed about the best fit. The first 50,000
elements of each walker were discarded in the ``burn-in'' phase during which
the chain reaches its stationary state. The typical autocorrelation length,
which is the interval over which the chain forgets its previous location, was
several thousand elements; the corresponding net number of independent samples
of the parameter space was $\sim 10^4$.  From the full distribution, we
trivially obtain a probability distribution for any given set of parameters of
interest by marginalizing over all the parameters that were outside that set.
Flat priors were adopted for all model parameters.  

\subsection{The four intrinsic parameters of the system}

We first discuss four parameters that are global for GX~339--4 (i.e., the same
for all six spectra), namely, the Galactic hydrogen column density
$N_\mathrm{H}$, the spin parameter $a_*$ of the black hole, the inclination
$i$, and the Fe abundance $A_\mathrm{Fe}$. The entries for these (and all
other) parameters estimated in the MCMC analysis given in Tables~\ref{tab:fit1}
and \ref{tab:fit2} are 90\% minimum-width confidence intervals about the
posterior maxima.

\subsubsection{Hydrogen column density}\label{sec:nh}

Despite the limited low-energy coverage of the PCA, the hydrogen column
density is well-constrained in both {\sc jf-i} and {\sc jf-ii} to
$N_\mathrm{H}=(6.5^{+0.8}_{-1.5}) \times 10^{21}$~cm$^{-2}$
(Tables~\ref{tab:fit1} and \ref{tab:fit2}), which is consistent with
other estimates in the literature, including (in units of
$10^{21}$~cm$^{-2}$) 4--6 \citep{kon00}; 6 \citep{zdz04}; 5.4
\citep{shi11}; and 5--8 \citep{men97}.

\subsubsection{Inclination of the inner disk}\label{sec:incl}

Previous estimates of inclination, which have been obtained by modeling the
reflected component, run the gamut (see Table~\ref{tab:master}). Three papers
using the same \xmm\ and \rxte\ data uniformly report low values:
\cite{mil06} and \cite{rei08} obtained $i=20^{+5}_{-15}$~deg and
$i\lesssim20$~deg, respectively, while \cite{don10}, using a different strategy
for reducing the data, found $i\sim 20-27$~deg. A much larger inclination,
$i=46\pm8$~degrees, was determined by \cite{shi11} using \suz\ data. All of
these results were obtained using the {\tt reflionx} models \citep{ros05}.
Recently, \cite{pla15} fitted simultaneously \xmm\ and \suz\ data sets using
both {\tt reflionx} and \xillver\ and reported two estimates of inclination:
$i=36^{+3}_{-6}$~deg and $i=42^{+11}_{-6}$~deg. In a different work, \cite{pla14b}
analyzed three low/hard state observations of GX~339--4 using a recent version
of the \relxill\ model and found $i=30^{+5}_{-4}$~deg, a result that may be
biased because it relies solely on low-energy \xmm\ data.

In {\sc jf-i}, we obtained a tight constraint on inclination, $i=48.4\pm 1.1$
deg (Table~\ref{tab:fit1}), a result that is consistent with that obtained in
{\sc jf-ii} (Table~\ref{tab:fit2}). Note that the \relxill\ and \xillver\ models used
here properly treat the angular distribution of the reflected radiation, unlike
earlier reflection models, which only provided an angle-averaged solution
\citep{gar14a}.

As an aside, if one makes the usual assumption that the spin of the black hole
is aligned with the orbital angular momentum vector \citep{fra10,ste12}, then
the lower estimates of inclination discussed above imply implausibly large
values of black hole mass based on the \citet{hyn03} estimate of the mass
function: e.g., $M \sim 5.8/\sin^3{i} \sim 100~M_{\odot}$ for $i\sim25$~deg.
Meanwhile, our inclination implies $M \sim 15M_{\odot}$, consistent with the
range of values observed for stellar-mass black holes \citep{oze10,far11}.

\begin{figure}
\centering
\includegraphics[scale=0.55]{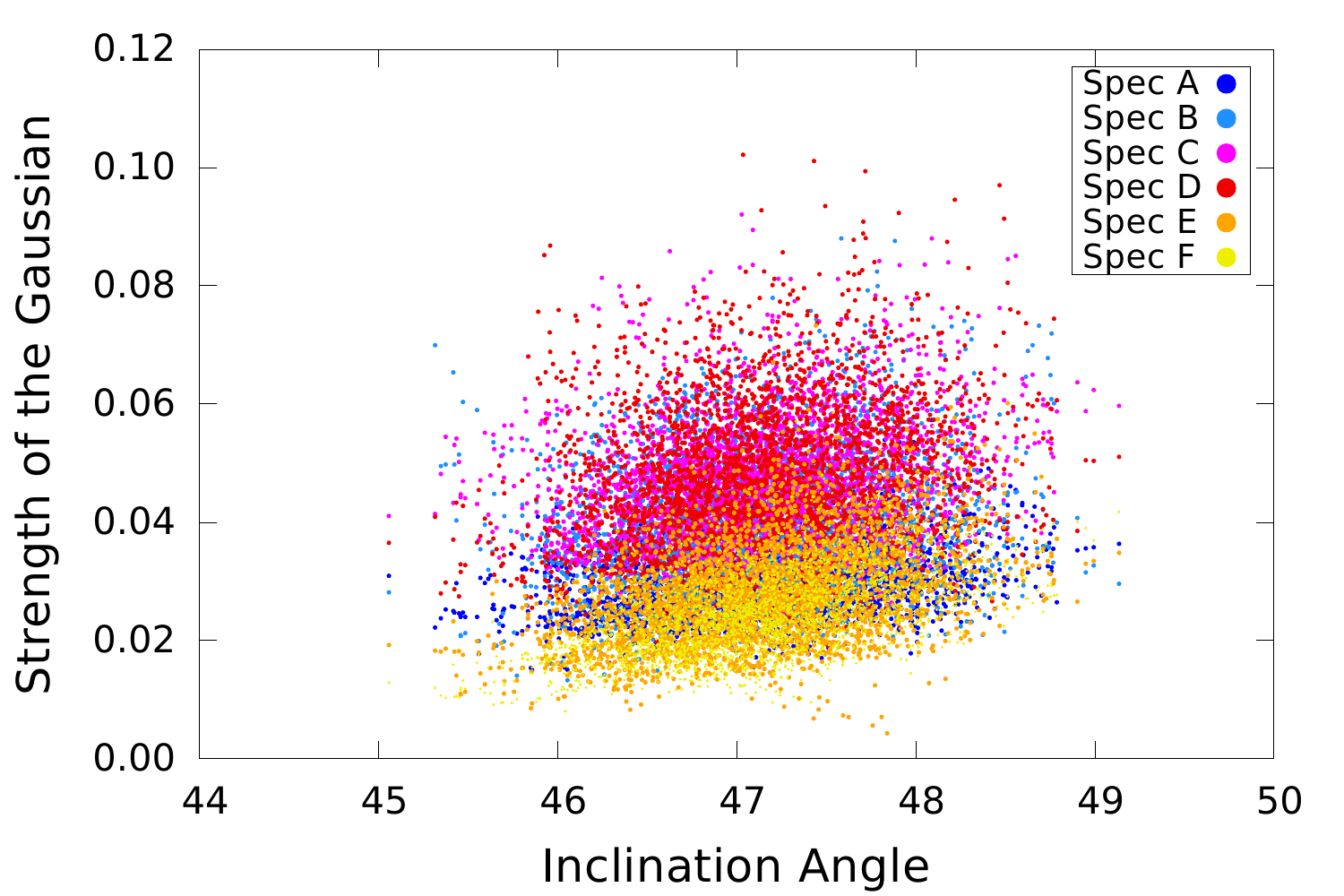}
\caption{
A probability density map for the strength of the Gaussian absorption
component ({\tt gabs}) and the inclination angle from our MCMC analysis
for {\sc jf-ii}. While the inclination is tied between the six spectra A--F, the
strength of the {\tt gabs} component is free to vary. The distribution
for each spectrum is coded by color. Only weak correlation is observed,
with the {\tt gabs} component becoming slightly stronger as the
inclination increases and as the luminosity decreases.
}
\label{fig:inclination}
\end{figure}

The inclination angle is largely determined by the shape and position of
the blue wing of the Fe K line, which in our fits is somewhat affected
by the inclusion -- or exclusion -- of the Gaussian absorption
feature. The energy of this feature was linked in all six spectra and
constrained to $7.23\pm 0.08$~keV (Table~\ref{tab:fit2}), while the
normalization was free to vary.  Figure~\ref{fig:inclination}, based on
our MCMC analysis, shows that while the strength of the {\tt gabs}
component does increase by a factor $\sim2$ with decreasing luminosity,
it only weakly interplays with inclination. This implies that this
component, whose origin is uncertain (Section~\ref{sec:gabs}), has at
most a modest affect on our estimate of inclination.

%
%

\subsubsection{Black hole spin}\label{sec:spin}

The spin of the black hole has been estimated via reflection modeling
using three independent data sets: $a_*=0.939\pm 0.004$ using
\xmm/EPIC-MOS plus \rxte\ spectral hard-state data \citep{mil06,rei08};
$a_*=0.93\pm 0.02$ using \xmm/EPIC-pn plus \rxte\ spectral data in the
very high (or steep power-law) state \citep{mil04,rei08}; and
$a_*=0.93\pm0.01$ (statistical) $\pm 0.04$ (systematic) using \suz\ data
\citep{mil08}. The corresponding estimates of inclination were all low
($i\sim 10-20$ deg) implying implausibly high estimates of black hole
mass assuming spin-orbit alignment (Section~\ref{sec:incl}).

There is considerable uncertainty associated with these estimates of
spin and inclination because of the effects of pileup, i.e., the arrival
of two or more photons in the same or adjacent CCD pixel within a single
frame time. For example, \cite{don10}, analyzing precisely the same
hard-state \xmm/EPIC-MOS data as \cite{mil06} and \cite{rei08}, conclude
that the high spin reported by Miller et al.\ and Reis et al.\ is the
result of severe pileup effects; using PN Timing-Mode data (presumably
unaffected by pileup), \cite{don10} report evidence for a narrow Fe line
and a truncated disk. Meantime, \cite{mil10} rebut the conclusions of
Done and D\'iaz-Trigo. As a second example, the high spin reported by
\citep{mil08} based on their analysis of \suz\ data (see above), is
challenged by \cite{yam09} who find -- using the same data set --
evidence for a truncated disk and no need to invoke a rapidly spinning
black hole.

\cite{kol10} applied the alternative continuum-fitting method to disk-dominated
\rxte\ data collected during three different outbursts of GX~339--4. This
method relies on accurate knowledge of the mass, distance, and inclination of
the system, all of which are highly uncertain for GX~339--4. Using approximate
bounds on these parameters, Kolehmainen \& Done obtained ``a strict upper
limit'' on the spin of $a_*<0.9$, which they claim is inconsistent with the
spin estimates obtained by modeling the reflection spectrum.

A chief virtue of the PCA data upon which we rely is its freedom from
the confusing effects of pileup. A further virtue is the abundance of
data, which allows us to track the behavior of GX~339--4 over a range of
luminosity, as well as to reach extreme ($\sim0.1$\%) levels of
statistical precision. By assuming that the inner radius of the disk
always remains at the ISCO ({\sc jf-ii}; Section~\ref{sec:fit2}), we
established a firm constraint on the spin at $a_* =
0.95^{+0.03}_{-0.05}$ (90\% confidence) while obtaining a precise
estimate of the inclination, $i=47.8^{+0.9}_{-1.4}$~degrees. Our spin
result is in accord with the earlier Fe-line estimates, but our
inclination estimate is distinctly different, and more in line with
expectation (Section~\ref{sec:incl}).

\begin{figure*}
\centering
\includegraphics[scale=0.9,angle=0]{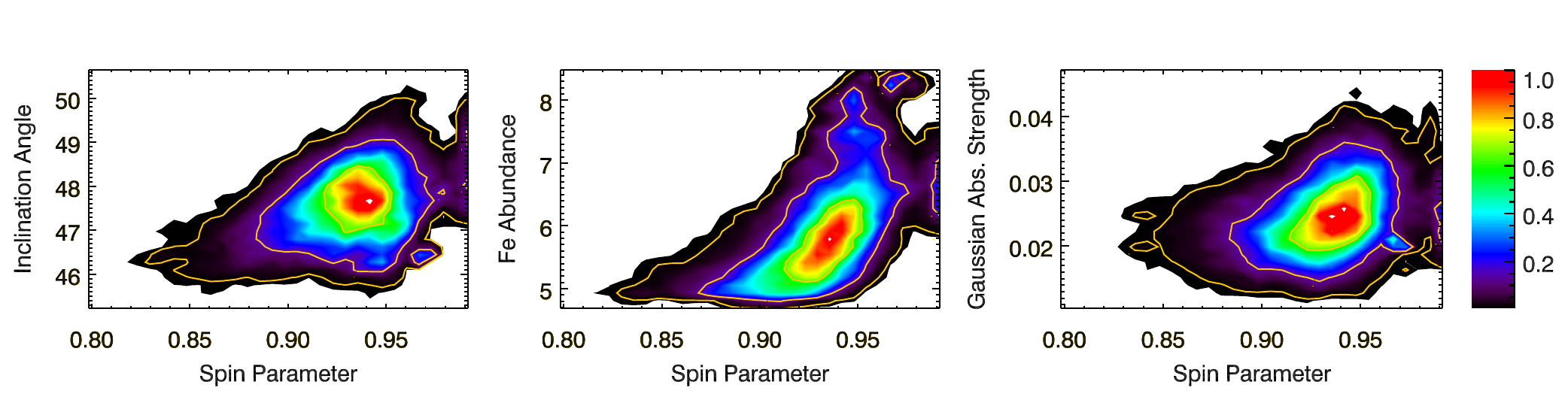}
\caption{ Probability contours from our MCMC analysis of {\sc jf-ii} for
  the spin parameter and three other parameters of interest: inclination
  angle ({\it left}), Fe abundance ({\it middle}), and the strength of
  the Gaussian absorption component ({\tt gabs}) for Spectrum A ({\it
    right}). A modest positive correlation is observed between the Fe
  abundance and the spin parameter.  }
\label{fig:spin}
\end{figure*}

The results of our MCMC analysis allow us to search for possible
degeneracies of the spin parameter with other fit parameters.
Figure~\ref{fig:spin} shows for {\sc jf-ii} spin probability
distributions for three key parameters: (1) the inclination angle, which
affects the blue wing of the line; (2) the Fe abundance, which affects
the strength of both the line and edge; and (3) the strength of the
Gaussian absorption, which also could affect the blue wing of the
line. While there is no evidence for a substantial correlation between
the spin and the inclination or the strength of the Gaussian, there is
indication of a moderate positive correlation with the Fe abundance.
From this positive correlation, it follows that an increase in Fe
abundance will produce a diminished inner radius for fits performed with
a fixed spin and variable $R_\mathrm{in}$ (Section~\ref{sec:rin}).

%
%

\subsubsection{Fe abundance}\label{sec:afe}

The Fe abundance for the blurred reflection component (\relxill) is
surprisingly high: $4.6^{+0.5}_{-0.3}$ and $5.4^{+1.9}_{-0.5}$ in solar
units for {\sc jf-i} and {\sc jf-ii}, respectively
(Tables~\ref{tab:fit1} and \ref{tab:fit2}).  Most studies have merely
assumed that the abundance is solar, while \cite{all13} fitted
\swi/\rxte\ hard-state data and also found a super-solar abundance:
$A_\mathrm{Fe}=2.4^{+1.47}_{-0.62}$. The high abundance results directly
from the remarkable strength of the Fe K line/edge relative to the
Compton hump (top panel of Figure~\ref{fig:fit1}), which is usually the
highest-amplitude feature in the reflection spectrum. A lower Fe
abundance underpredicts the strength of the line/edge required to fit
the Compton hump. Thus, our ability to constrain the Fe abundance is
likely a consequence of our broad bandpass that provides
high-sensitivity coverage of all the principal reflection features, from
the Fe K line on through complete coverage of the Compton hump. This
quality of coverage is not provided by \xmm\ data (even when \rxte\ data
with a floating normalization are included) which may explain why others
have not reported a super-solar Fe abundance.

\begin{figure}[hb]
\centering
\includegraphics[scale=0.4,trim={0 2cm 0 0}]{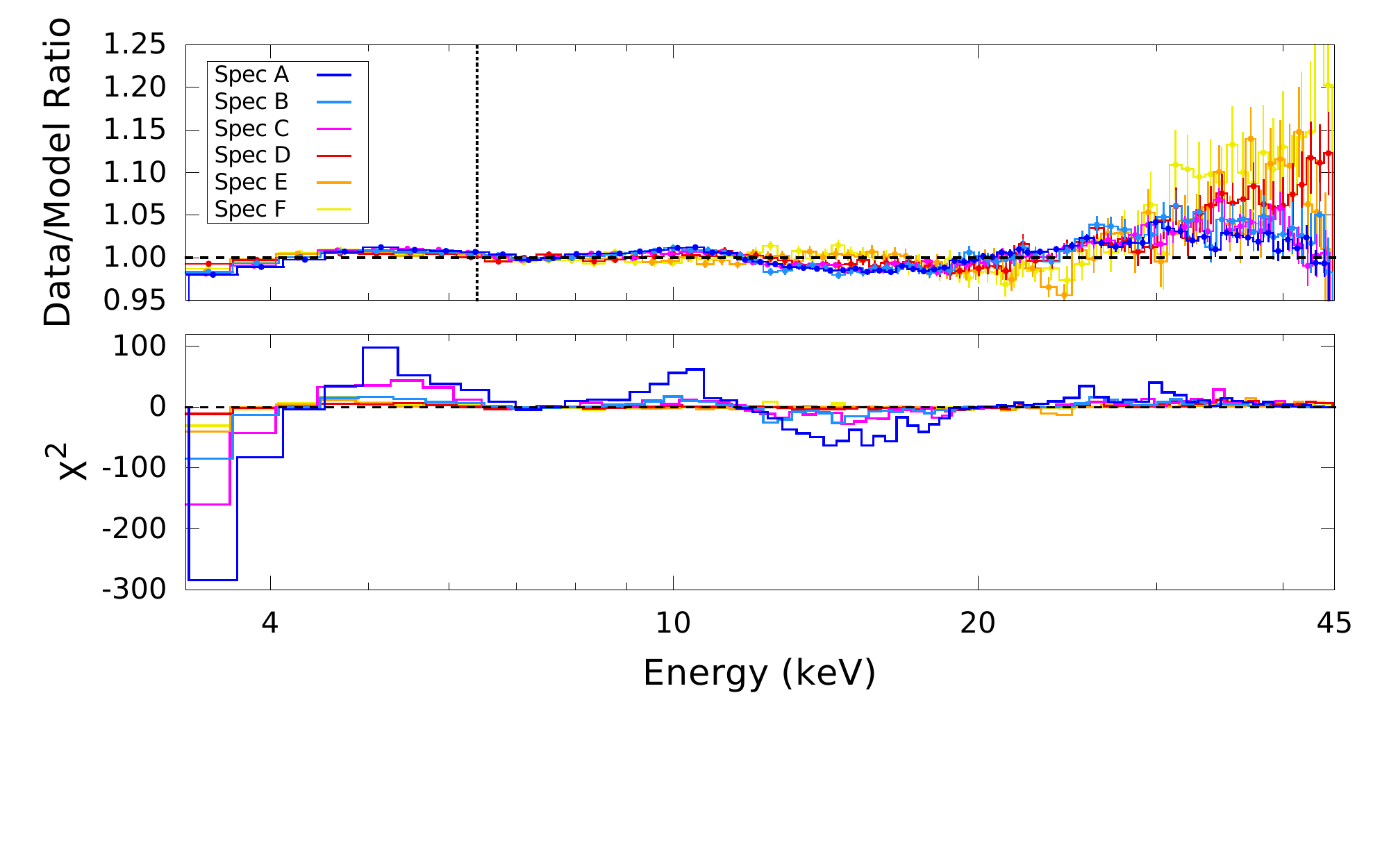}
\caption{Spectral fitting results, similar to {\sc jf-i}, but here fixing the
  Fe abundance to the solar value ($A_\mathrm{Fe}=1$).  The top and
  bottom panels show respectively a ratio plot and contributions to
  $\chi^2$.  This model fails to fit jointly the three principal
  reflection features (Fe K line and edge and the Compton hump), most
  noticeably for the higher-luminosity spectra, and the fit is
  unacceptable ($\chi^2_{\nu}=9.9$).  A vertical dashed line marks the
  6.4~keV rest-frame energy of the Fe K line.  }
\label{fig:afe-1}
\end{figure}

Quantitatively, forcing the Fe abundance to the solar value ($A_\mathrm{Fe}=1$)
results in a grossly unacceptable fit with $\chi^2_{\nu}=9.9$ and large
residuals across the PCA band (Figure~\ref{fig:afe-1}), while the inner-disk
radius grows by about a factor of 10 compared to the fit with variable Fe
abundance.  We tried several alternative models (e.g., varying the emissivity
index) in an unsuccessful attempt to find an acceptable model with lower Fe
abundance. We note that fits to \nustar\ data for the most recent outburst of
GX~339--4 likewise require a large Fe abundance \citep{fue15}.

%
%

The Fe abundance has an influence on the shape of the Fe K line, and it may
therefore in turn affect such parameters of interest as the inclination or
inner-disk radius. Moreover, the abundance also alters the continuum
photoelectric opacity at higher energies, which modifies the depth of the Fe K
edge and the red side of the Compton hump. These effects, which are subtle, are
driving our fits because of the unprecedented signal-to-noise we have achieved.
We shall return to this point in Section~\ref{sec:rin}.

We conclude that super-solar Fe abundance is a strong and inescapable
requirement of these data. At the same time, the data also require the
unblurred (distant) reflection to have moderate -- near-solar -- Fe
abundance.  In our tests, fitting the spectrum with the most counts
(Spectrum~A), we found that adopting a single abundance for both blurred
and unblurred components (i.e., \relxill\ and \xillver) results in
$A_\mathrm{Fe}\sim 3.6$, an intermediate value between \relxill\
($A_\mathrm{Fe}\sim 5$), and the low value required for the \xillver\
component.  Though this approach may be intuitively more satisfying, it
is strongly rejected by the data, with an increase in $\chi^2$ of $\sim
55$.  The effect of linking the abundances is to decrease the importance
of the \xillver\ component by reducing its normalization parameter from
$\sim 0.2$ to $\sim 0.07$.  An inspection of the residual contributions
to $\chi^2$ (lower panel of Figure~\ref{fig:afe-1}) reveals that the
quality of the fit is degraded not only in the Fe K region but over the
entire energy band. In summary, we find strong empirical evidence for
the presence of an unblurred reflection component whose Fe abundance is
much less than that of the blurred component.

At the same time, there is no obvious reason why the inner-disk abundances
should be so high. We note that similar physical processes may be occurring in
AGN, since large Fe abundances are likewise found in many cases when fitting
relativistic reflection models \citep{fab06}, with 1H~0707--495 being a prime
example \citep{dau12,kar15b}. Possibly, an unknown physical effect is being
overlooked in current models that is artificially driving the Fe abundance to
high values.  For example, \cite{rey12} proposed that radiative levitation of
Fe ions in the accretion disk atmosphere could cause an apparent enhancement of
their abundance. 

\subsection{Parameters that evolve systematically with luminosity}

Setting aside the Gaussian absorption component (already discussed in
Section~\ref{sec:fitcomb}), for {\sc jf-i} there are six important parameters
that are fitted separately for Spectra~A--F: the inner-disk radius
$R_{\rm in}$; the photon index $\Gamma$; the ionization parameter $\xi$; the
high-energy cutoff $E_{\rm cut}$; the reflection fraction $R_{\rm f}$; and the
normalization of the unblurred reflection component (\xillver). In this
section, we show how these parameters depend on luminosity, and we discuss the
causes of these dependencies.

\begin{figure*}
\centering
\includegraphics[scale=0.70,trim={0 4.0cm 0 0}]{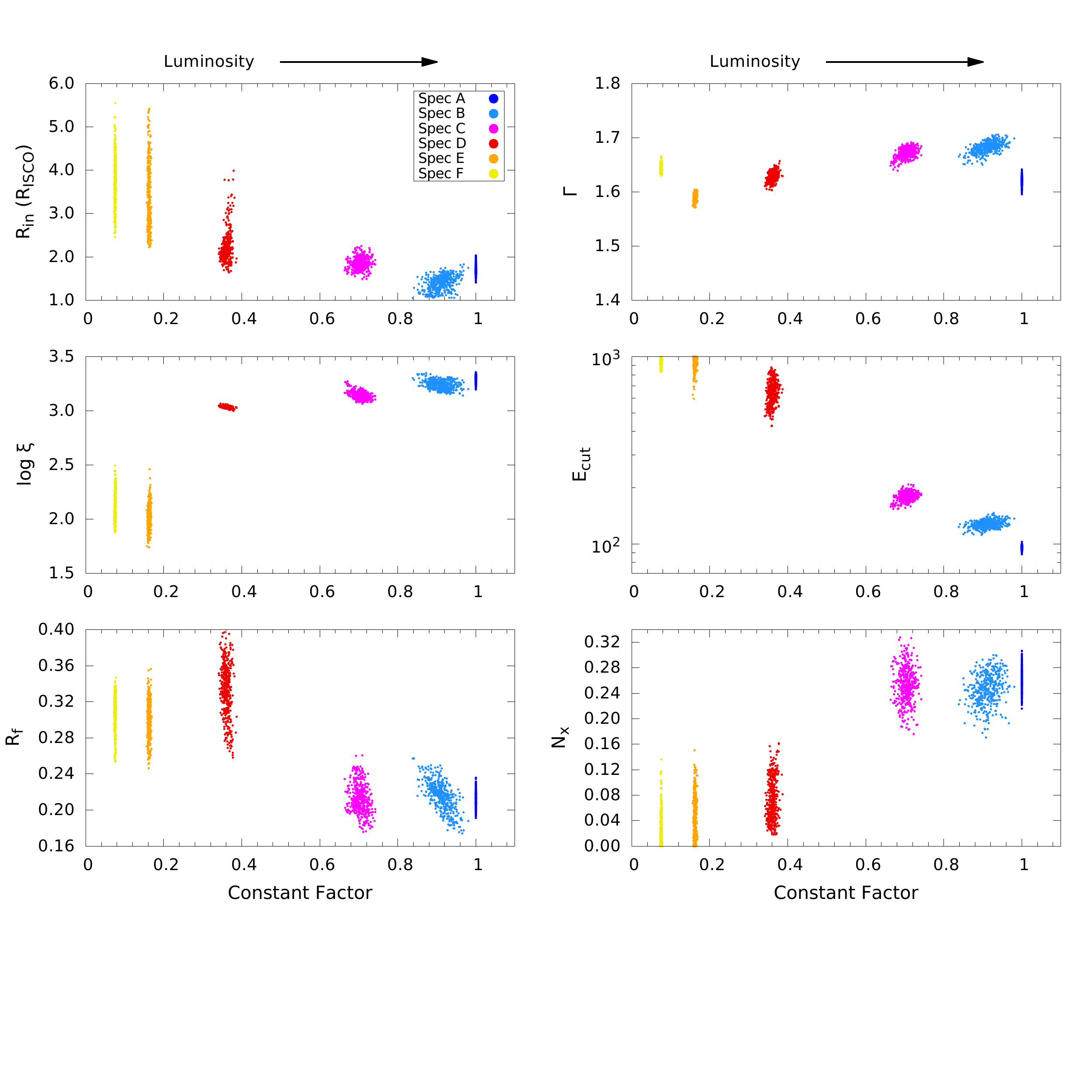}
\caption{
Variation of key model parameters with X-ray luminosity. The clouds of
points in each panel (color-coded to correspond to a particular one of
the six spectra) show the posterior density of the MCMC results for: the
inner radius $R_\mathrm{in}$ in units of the ISCO radius; the photon
index $\Gamma$ of the power law; the ionization parameter $\xi$; the
high-energy cutoff $E_\mathrm{cut}$; the reflection fraction $R_\mathrm{f}$; and
the normalization $N_\mathrm{x}$ of the unblurred reflection component
\xillver.  The Constant Factor on the $x$-axis, which is proportional to
the Eddington-scaled luminosity, is normalized to unity (corresponding
to $L/L_{\rm Edd} = 17$~\%) for Spectrum A (Table~\ref{tab:boxes}).
}
\label{fig:params}
\end{figure*}

Figure~\ref{fig:params} illustrates our MCMC results for {\sc jf-i}
(Table~\ref{tab:fit1}), the case of fixed spin. The probability
distribution for each parameter is shown plotted versus the floating
constant factor, which can be regarded as a proxy for the
luminosity. The luminosity ranges over somewhat more than an order of
magnitude. Each Spectrum is color-coded (see legend in top-left
panel). The breadth of a distribution is a measure of uncertainty, while
its shape indicates the degree of correlation of that particular
parameter with luminosity. We now discuss in turn the behavior of each
parameter.

%
%
%

%
\begin{figure*}
\centering
\includegraphics[scale=0.6]{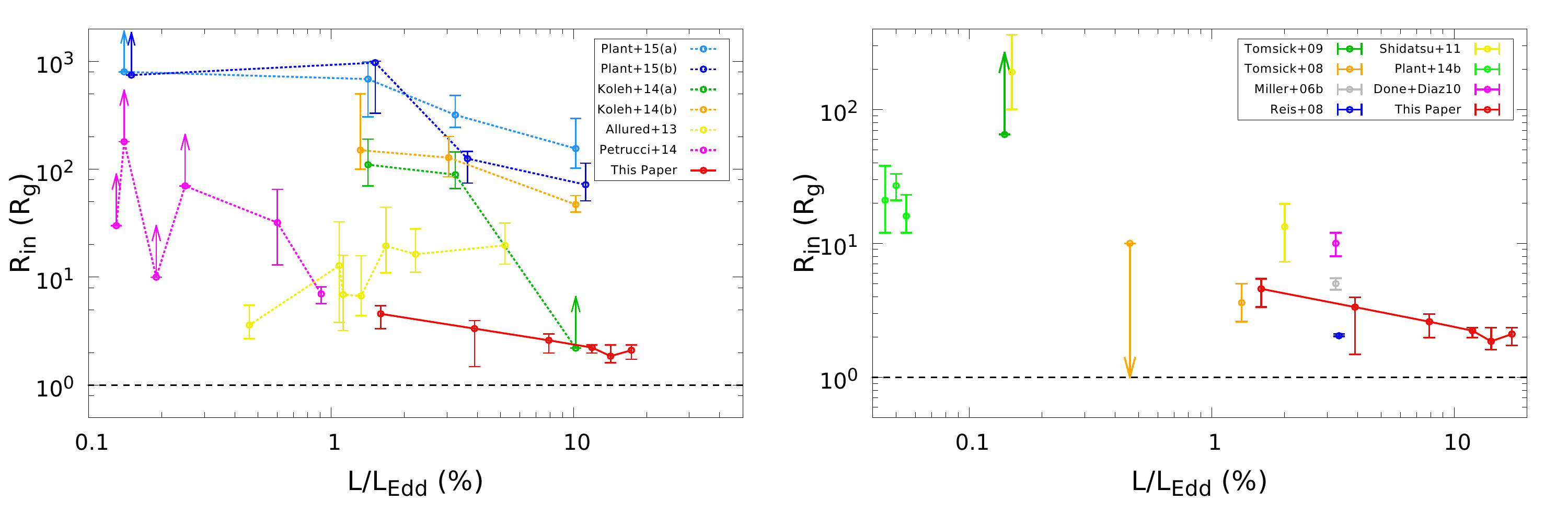}
\caption{ A comparison for GX~339--4 of our estimates with those in the
  literature (see Table~\ref{tab:master}) of the inner-disk radius
  versus luminosity obtained by reflection modeling of hard-state
  spectra. The solid red track shown in both panels links our six
  measurements of $R_\mathrm{in}$ (Table~\ref{tab:fit1}). ({\it left})
  The various colored tracks show the evolution of $R_\mathrm{in}$ with
  Eddington-scaled luminosity (see footnote to Table~\ref{tab:boxes})
  reported by us and others (Table~\ref{tab:master}).  \cite{pla15} and
  \cite{kol14} each present a pair of tracks, which are labeled {\tt
    (a)} and {\tt (b)}; the latter track in the case of \cite{kol14}
  includes a notch feature at $\sim 9$~keV. ({\it right}) Single
  measurements of $R_\mathrm{in}$ reported in a variety of studies
  (Table~\ref{tab:master}).  The track labeled {\tt Done+Diaz10}
  corresponds to results reported by \cite{don10} for fits to Epic-pn
  data. Data points with identical values of luminosity are slightly
  offset for clarity (precise values are given in
  Table~\ref{tab:master}). }
\label{fig:rin}
\end{figure*}
%

%
\begin{table*}
\caption{Compilation of literature estimates of $R_\mathrm{in}$ obtained
by fitting reflection models to hard-state spectra of GX~339--4.
}
\begin{center}
\label{tab:master}
\footnotesize
\begin{tabular}{llcccccc}
\hline
Satellite & Instrument & $L/L_\mathrm{Edd}$~(\%) & $R_\mathrm{in}$ ($R_g$) & $i$ (deg) & $q$ & High-energy? & Ref. \\
\hline
\xmm\ & EPIC-pn (TM)\tablenotemark{a} & 1.42    & $684^{+301}_{-378}$   & $42^{+11}_{-6}$ & 3             & Yes   & [1] \\
      &              &         & $972^{+28}_{-643}$    & $36^{+3}_{-6}$  & 3             & Yes   & [2] \\
      &              &         & $110^{+80}_{-40}$     & 60              & 3             & No    & [3] \\
      &              &         & $150^{*}_{-50}$       & 60              & 3             & No    & [4] \\
      & EPIC-MOS     & 3.25    & $5\pm 0.5$              & $20^{+5}_{-10}$ & 3             & Yes   & [5] \\
      &              &         & $2.04^{+0.07}_{-0.02}$& $20_{-1.3}$     & $3.16\pm 0.5$ & Yes   & [6] \\
      & EPIC-pn      &         & $10\pm 2$             & $27\pm 3$       & 3             & Yes   & [7] \\
      &              &         & $60^{+40}_{-20}$      & 60              & 3             & Yes   & [8] \\
      & EPIC-pn (TM)\tablenotemark{a} &         & $318^{+165}_{-74}$    & $42^{+11}_{-6}$ & 3             & Yes   & [1] \\
      &              &         & $125^{+21}_{-51}$     & $36^{+3}_{-6}$  & 3             & Yes   & [2] \\
      &              &         & $89^{+55}_{-23}$      & 60              & 3             & No    & [3] \\
      &              &         & $128^{+73}_{-43}$     & 60              & 3             & No    & [4] \\
      &              & 10.2    & $155^{+139}_{-53}$    & $42^{+11}_{-6}$ & 3             & Yes   & [1] \\
      &              &         & $72^{+42}_{-21}$      & $36^{+3}_{-6}$  & 3             & Yes   & [2] \\
      &              &         & $2.2^*$               & 60              & 3             & No    & [3] \\
      &              &         & $47^{+10}_{-7}$       & 60              & 3             & No    & [4] \\
      &              & $<0.05$ & $21^{+17}_{-9}$       & $30^{+5}_{-4}$  & 3             & No    & [14] \\
      &              & $<0.05$ & $27^{+6}_{-6}$        & $30^{+5}_{-4}$  & 3             & No    & [14] \\
      &              & $<0.05$ & $16^{+7}_{-4}$        & $30^{+5}_{-4}$  & 3             & No    & [14] \\
\suz\ & XIS0,1,3/PIN & 0.14    & $>65$                 & 18              & 2--3          & Yes   & [9] \\
      &              &         & $>798$                & $42^{+11}_{-6}$ & 3             & Yes   & [1] \\
      &              &         & $>745$                & $36^{+3}_{-6}$  & 3             & Yes   & [2] \\
      &              &         & $190^{+170}_{-90}$    & 50              & 2.3           & Yes   & [10] \\
      &              &         & $>180$                & 20              & 3             & Yes   & [11] \\
      &              & 0.13    & $>30$                 & 20              & 3             & Yes   & [11] \\
      &              & 0.19    & $>10$                 & 20              & 3             & Yes   & [11] \\
      &              & 0.25    & $>70$                 & 20              & 3             & Yes   & [11] \\
      &              & 0.60    & $32^{+33}_{-19}$      & 20              & 3             & Yes   & [11] \\
      &              & 0.91    & $7.0^{+1.1}_{-1.3}$   & 20              & 3             & Yes   & [11] \\
      &              & 2.0     & $13.3^{+6.4}_{-6.0}$  & $46\pm 8$       & $2.3\pm 0.1$  & Yes   & [10] \\
\swi  & XRT          & 1.33    & $3.6^{+1.4}_{-1.0}$   & 20              & $3.2\pm 0.6$  & Yes   & [12] \\
      &              &         & $6.7^{+9.1}_{-2.3}$   & 20              & 3             & Yes   & [13] \\
      &              & 0.46    & $<10$                 & 20              & $3.1\pm 0.4$  & Yes   & [12] \\
      &              &         & $3.6^{+1.9}_{-0.9}$   & 20              & 3             & Yes   & [13] \\
      &              & 1.08    & $12.8^{+19.8}_{-9.0}$ & 20              & 3             & Yes   & [13] \\
      &              & 1.12    & $6.9^{+9.1}_{-3.7}$   & 20              & 3             & Yes   & [13] \\
      &              & 1.68    & $19.5^{+25}_{-8.5}$   & 20              & 3             & Yes   & [13] \\
      &              & 2.23    & $16.3^{+11.7}_{-5.2}$ & 20              & 3             & Yes   & [13] \\
      &              & 5.21    & $19.7^{+12.1}_{-6.5}$ & 20              & 3             & Yes   & [13] \\
\hline
\end{tabular}
\end{center}
$^a$ {TM=Timing Mode}
\tablecomments{[1] \cite{pla15} implementing \xillver;
  [2] \cite{pla15} implementing {\tt reflionx}; [3] \cite{kol14}; [4]
  \cite{kol14} including a notch feature at $\sim 9$~keV; [5] \cite{mil06}; [6]
  \cite{rei08}; [7] \cite{don10}; [8] \cite{don10} with fixed inclination; [9]
  \cite{tom09}; [10] \cite{shi11}; [11] \cite{pet14}; [12] \cite{tom08};
  [13] \cite{all13}; [14] \cite{pla14b}.  }
\end{table*}

\subsection{Inner edge of the disk}\label{sec:rin}

The evolution of the inner-disk radius $R_\mathrm{in}$ with luminosity is shown
in the top-left panel of Figure~\ref{fig:params}. Each spectrum delivers a
good constraint on $R_\mathrm{in}$, allowing us to conclude that the inner edge
of the disk moves outward by a factor of a few as the luminosity decreases by an
order of magnitude, from a nominal value of 17\% of Eddington to 1.6\% of
Eddington (Table~\ref{tab:boxes}).

This is a principal result of our paper because $R_\mathrm{in}$ and its
dependence on luminosity is a matter of central importance for the study
of black hole binaries in the hard state
(Section~\ref{sec:controversy}). In Table~\ref{tab:master} we summarize
estimates of $R_\mathrm{in}$ in the literature for GX~339--4\ in the
hard state, while considering only those results obtained via reflection
spectroscopy. The compilation includes results obtained using a wide
variety of data and over a large range in luminosity ($\sim0.1-20$\% of
Eddington). At a glance, one notes the extreme range of values reported
for $R_\mathrm{in}$. The most notable conflict are two grossly disparate
values reported for the same \xmm\ observation: \cite{rei08} analyzed
MOS and \rxte\ PCA data and reported
$R_\mathrm{in}=2.04^{+0.07}_{-0.02}$~$R_g$, while \cite{pla15} analyzed
EPIC-pn timing-mode data and reported
$R_\mathrm{in}=318^{+165}_{-74}$~$R_g$.

Figure~\ref{fig:rin} shows all the values of $R_\mathrm{in}$ that appear
in Table~\ref{tab:master} plotted as a function of the Eddington-scaled
luminosity. Several studies report results for multiple observations
over a range of luminosity; in these cases, the individual data points
are highlighted in the left panel of Figure~\ref{fig:rin} using colored
tracks. Meanwhile, individual measurements are shown in the right
panel. Our results are shown in both panels with a solid red track
connecting the data points. As noted above, we find that $R_\mathrm{in}$
increases modestly with decreasing luminosity: Best-fit values trend
upward from $2.1~R_g$ to $4.6~R_g$ as the luminosity decreases from
17\% to 1.6\% of Eddington. (Note that the values of $R_\mathrm{in}$
in Table~\ref{tab:fit1} are in units of $R_\mathrm{ISCO}=1.237R_g$.)

This trend is consistent with that found in previous studies except for
that of \cite{all13} (yellow track), who fitted \swi\ and \rxte\ data
using the {\tt reflionx} model.  Despite the general agreement that the
inner radius shrinks with increasing luminosity, our estimates of
$R_\mathrm{in}$ at comparable values of luminosity are much smaller than
those reported by others.  

For example, \cite{pla15} found the disk to be extremely truncated based
on fits to a \suz\ and three \xmm\/\rxte\ spectra using \xillver\ and
{\tt reflionx} (light and dark blue tracks). \cite{kol14} reported a
similar trend but smaller values of radius (green track) by analyzing
the same three \xmm\ spectra (excluding the \rxte\ data) using the {\tt
  rfxconv} model \citep[based on the {\tt reflionx} tables;][]{kol11}.
For their highest-luminosity data, they report two values of inner
radius: One is a lower limit that is consistent with our results,
$R_\mathrm{in} > 2.2 R_g$, while the other (which includes a $\sim
9$~keV instrumental feature in the fit; orange track) is reasonably
consistent with the results of \cite{pla15}, $R_\mathrm{in} \approx
47~R_g$.  Overall, the results of \cite{kol14} and \cite{pla15} are
similar, with the former authors reporting somewhat smaller values of
$R_\mathrm{in}$.  Interestingly, our results appear to be in reasonable
agreement with an extrapolation of the low-luminosity values of
$R_\mathrm{in}$ reported by \cite{pet14} (purple track), which is not
necessarily expected since these authors assumed solar Fe abundance.

Though the gross disparities in the reported values of $R_\mathrm{in}$
may be partially due to differences in the models, this should be a
secondary effect since, e.g., tests show that the models \xillver\ and
{\tt reflionx} perform similarly \citep{gar13a}. The more likely reason
for the inconsistent results is limitations of the data. One of the most
severe of these is the effects of pileup, especially for the crucial
\xmm\ data (see Section~\ref{sec:spin}).  Another effect leading to
major differences in results is whether or not high-energy data were
used. For example, \cite{pla15} and \cite{kol14} used the same EPIC-pn
data in timing mode, but while \cite{pla15} used simultaneous \rxte\
data to extend the energy coverage, \cite{kol14} eschewed its use
because of their concern over the cross-calibration of the two
detectors. As a consequence of employing \xmm\ data only, the results of
\cite{kol14} are highly sensitive to calibration issues associated with
the rapidly-falling and uncertain response of the EPIC-pn detector at
energies $\gtrsim 9$~keV.



The PCA has important advantages despite its limited energy resolution
and lack of coverage below 3~keV. Most notably, the PCA data are free
from the contentious effects of pileup that are inherent to CCD
observations of bright sources. Meanwhile, the use of a single detector
eliminates problems associated with cross-calibrating a pair of
detectors. The much higher effective area of the PCA around the Fe line
and Compton hump -- and the many dozens of observations -- yields
spectra with orders-of-magnitude more counts than CCD spectra
(Section~\ref{sec:simultaneous}). Moreover, one can now fully utilize
these many millions of counts per spectrum to detect subtle effects in
reflection features because the response of the PCA has been
successfully calibrated to $\sim0.1$\% precision
\citep{sha12,gar14b}. These virtues of the PCA data are attested to by
our success in fitting our reflection models ($\chi^2_{\nu}\sim 1$) to
six extremely high signal-to-noise spectra, which individually contain
between 3 and 28 million total counts in the 3--10 keV band. Finally,
the great abundance of data makes the PCA database unrivaled for
synoptic studies of Galactic black holes.

\begin{figure}
\centering
\includegraphics[scale=0.55]{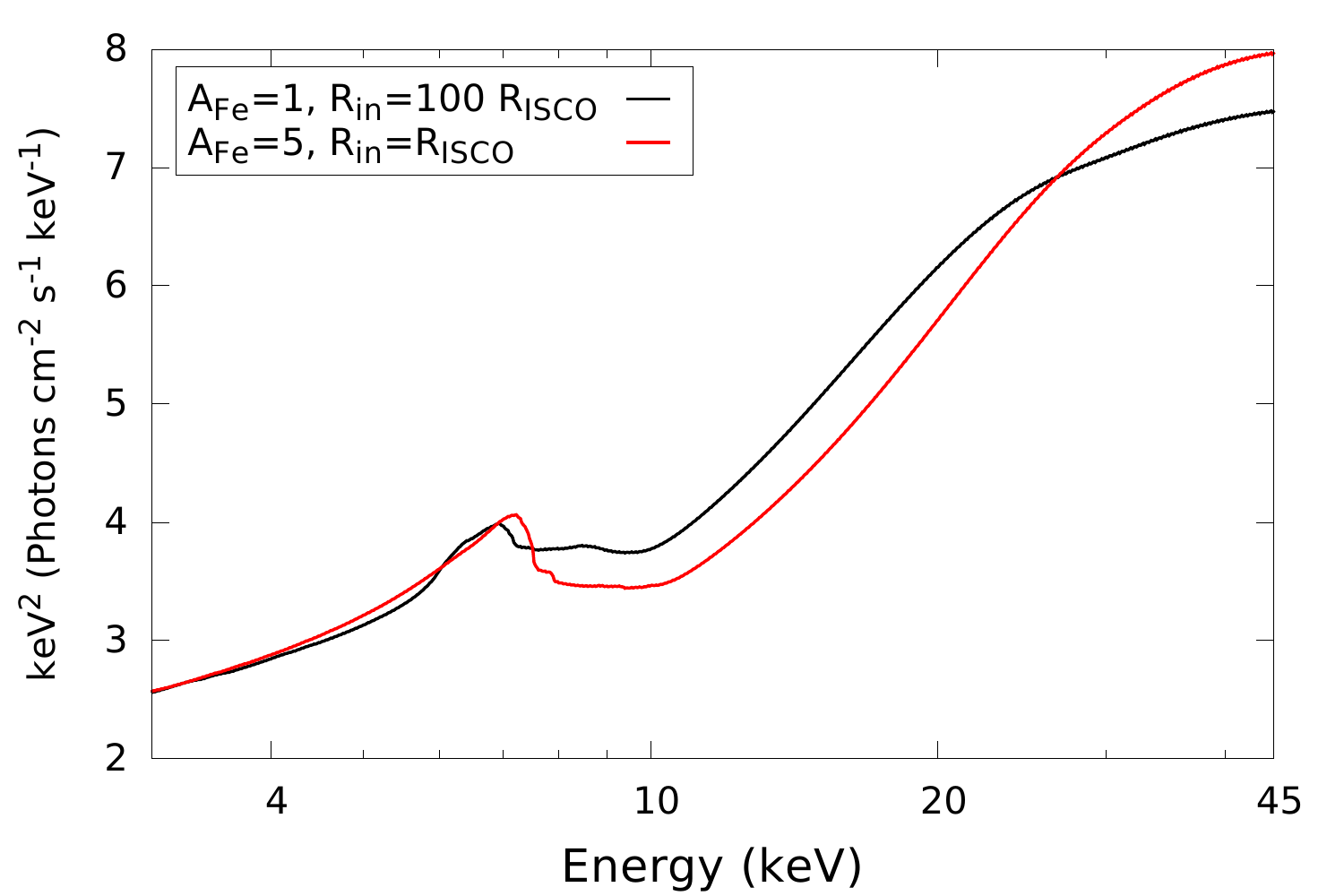}
\caption{
A comparison of two reflection models calculated using our \relxill\
code: The red curve is for the parameters of {\sc jf-i} (see
Table~\ref{tab:fit1}), in particular for $A_\mathrm{Fe}=5$ and
$R_\mathrm{in}=R_\mathrm{ISCO}$.  The black curve is for a model with
solar abundances ($A_\mathrm{Fe}=1$) that has been doctored by
increasing $R_\mathrm{ISCO}$ by 100-fold to best match the structure in
the Fe K line.  Meantime, the models are seen to differ greatly in the
depth and structure of the Fe K edge and in the shape of the Compton
hump.
}
\label{fig:rinmodel}
\end{figure}

We now return to the question of the grossly discrepant results reported
for $R_\mathrm{in}$ (Table~\ref{tab:master}; Figure~\ref{fig:rin}) while
reminding the reader that the Fe abundance affects the Fe K line profile
and other reflection features at a detectable level given our signal to
noise (Section~\ref{sec:afe}). In turn, the Fe abundance affects other
parameters, notably the inner-disk radius and spin parameter, which
correlates positively with Fe abundance (Section~\ref{sec:spin}). We now
show that values of $R_\mathrm{in}$ found by others are significantly
biased by either the low signal-to-noise of their data or inadequate
high-energy coverage. Such data make it difficult to distinguish between
small $R_\mathrm{in}$ with large $A_\mathrm{Fe}$ and large $R_{\rm in}$
with solar abundances, as we illustrate in Figure~\ref{fig:rinmodel},
which compares fits to two \relxill\ models, one with $A_\mathrm{Fe}=1$
and the other with $A_\mathrm{Fe}=5$. The model with solar abundance
(black curve) can only fit the data when the disk is strongly truncated,
which serves to minimize the relativistic effects that blur the line
profile. Note, however, that this model then fails to reproduce the
depth of the Fe K edge and underpredicts the continuum above
$\sim30$~keV.  Figure~\ref{fig:rinmodel} should be compared directly
with Figure~\ref{fig:afe-1}, where the limitations of the
$A_\mathrm{Fe}=1$ model are apparent from the fit residuals. Unlike most
other data sets, the extreme signal in our data clearly discriminates
between the two models.

The truncation of the inner disk and the decrease in $R_\mathrm{in}$ with
increasing $L/L_\mathrm{Edd}$, which we find, is a prediction of the
advection-dominated accretion flow (ADAF) model (see
Section~\ref{sec:controversy}). In this paradigm, the inner disk evaporates
becoming a very hot and optically thin accretion flow that fills the inner
region (see, e.g., \citealt{mey09}). Our results are in line with this model,
although our observations do not extend to the lower luminosities at which
extreme truncation likely occurs.

However, we note that these results are apparently at odds with our
non detection of a thermal disk component of emission. Making the
usual assumption that all the observed power-law photons are generated
by Compton up-scattering of disk photons, we would have expected to
detect a thermal component for a disk that extends to such small radii
(see Section~\ref{sec:conc}).


%
%

\subsection{Parameters of the continuum: $\Gamma$ and $E_\mathrm{cut}$}\label{sec:corona}

We find that the power-law photon index $\Gamma$ is relatively constant despite
the order of magnitude increase in luminosity (top-right panel of
Figure~\ref{fig:params}), a result that has been previously reported for
GX~339--4 \citep[e.g.][]{wil99,zdz04,pla14}. Its average value is
$1.640\pm0.035$ for {\sc jf-i} and $1.625\pm0.030$ for {\sc jf-ii} (std.\ dev., $N=6$;
Table~\ref{tab:fit1} and \ref{tab:fit2}), firmly in the range for the hard
state \citep[$1.4<\Gamma<2.1$;][]{rem06}.

In contrast to the constancy of the power-law index, the cutoff energy
$E_\mathrm{cut}$ systematically decreases with increasing luminosity
from $>890$~keV for Spectrum~F down to $97 \pm 4$~keV for Spectrum~A
(right-middle panel of Figure~\ref{fig:params}). This lower value of
$E_\mathrm{cut}$ for our highest luminosity spectrum is of the same
order of magnitude as the $58.5\pm 2.2$~keV value reported by
\cite{dro10}, which is based on their analysis of simultaneous \rxte\
and {\it INTEGRAL} data obtained during another bright hard state of
GX~339--4.

Our model achieves good constraints for all six spectra. Remarkably,
this is true for even the lowest-luminosity data (Spectrum F) for which
the cutoff energy of $>890$ keV~is far beyond the 45~keV limit of the
PCA bandpass.  This surprising result is a consequence of the detectable
effects that are imprinted on the reflected component in the 3--45~keV
band by photons with energies of hundreds of keV. We discuss the
capability of the \relxill\ model to probe the spectrum at extreme
energies in \citet{gar_ecut_2015}.

In a Comptonized and isothermal corona, the high-energy cutoff is set by the
electron temperature: $E_\mathrm{cut} \sim (2-3)kT_e$. In such a plasma,
thermal disk photons are Compton up-scattered, thereby cooling the coronal
electrons while producing the observed power-law continuum. The slope of the
power law depends on the interplay between the electron temperature and the
optical depth $\tau_e$,
\begin{equation}\label{eq:gamma}
\Gamma = -\frac{1}{2} + \sqrt{\frac{9}{4} + \frac{1}{\theta_e \tau_e (1+\tau_e/3)} },
\end{equation}
\citep{lig87}, where $\theta_e=kT_e/m_ec^2$ and $m_ec^2=511$~keV is the
electron rest mass. Values of these parameters for Spectra A--F, which are
consistent with previous determinations \citep[e.g.][]{wil99}, are summarized
in Table~\ref{tab:corona} for our nominal value of the photon index
($\Gamma=1.6$). We find, as predicted by Equation~\ref{eq:gamma}, that the
coronal temperature decreases with increasing luminosity, while the optical
depth increases.

%
\begin{table}
\caption{Coronal properties$^a$}
\begin{center}
\label{tab:corona}
\footnotesize
\begin{tabular}{ccccc}
\hline
Box & $L/L_\mathrm{Edd}$ & $E_\mathrm{cut}$ & $T_e$        & $\tau_e$  \\
    & (\%)               & (keV)            & ($10^9$~K) &           \\
\hline
A & 17.3 & 97 & 0.45 & 3.03 \\
B & 14.2 &129 & 0.60 & 2.50 \\
C & 11.9 &179 & 0.83 & 1.99 \\
D &  7.9 &660 & 3.06 & 0.72 \\
E &  3.9 &840 & 3.90 & 0.59 \\
F &  1.6 &890 & 4.13 & 0.56 \\
\hline
\end{tabular}
\end{center}
$^a$ Note. -- Assumes $kT = \frac{2}{5}E_\mathrm{cut}$ and $\Gamma=1.6$ 
(See Equation~\ref{eq:gamma}).
\end{table}
%

\subsection{The reflection fraction}\label{sec:refrac}

The reflection fraction $R_{\rm f}$ is a third parameter (in addition to
$\Gamma$ and $E_\mathrm{cut}$) that provides information on the
structure of the corona. In \relxill, the parameter is empirically
defined as the ratio of the reflected flux to the power-law flux in the
20--40~keV band.  The results of {\sc jf-i} show that the reflection
fraction of the relativistically blurred component (\relxill) ranges
from $0.2 \lesssim R_{\rm f} \lesssim 0.3$, decreasing modestly with
increasing luminosity (bottom-left panel, Figure~\ref{fig:params}). This
trend is surprising given that, at the same time, $R_{\rm in}$ is
decreasing so that the area of the reflector should increase. As a
further wrinkle, one expects $R_{\rm f} \gtrsim 1$ based on simple
arguments \citep{dau14}.

There are several scenarios that can plausibly account for values of $R_{\rm f}
< 1$. We mention four and then discuss a new, alternative explanation. (1) An
obvious explanation is a severely truncated disk (more specifically, a disk
with truncation radius large compared to the size of the corona). We discard
this possibility as inconsistent with the small values we find for $R_{\rm in}$
(Section~\ref{sec:rin}). (2) Another option is for the corona to be
continuously outflowing at relativistic speeds, beaming the bulk of its
emission away from the disk \citep[e.g.][]{mil14,kec15}. While this is a
possible explanation for $R_{\rm f} < 1$, one outcome of this scenario is a
resultant low value of the emissivity index, which is not obviously required by
our data.  (3) The value of $R_{\rm f}$ may be depressed by our assumption of a
constant-density disk atmosphere, as \cite{bal01} have shown in their studies
of hydrostatic atmospheres. The hotter gas layer at the surface of a
hydrostatic atmosphere additionally scatters and blurs reflection features
\citep[see also][]{nay01} thereby diluting the reflection signal relative to a
constant density model. (4) The apparent strength of reflection features may
also be reduced by the Comptonization of these features in an extended
corona, as \cite{wil14} recently proposed. However, for such a corona to be
effective in reducing $R_{\rm f}$ appreciably, it must have a large covering
fraction which may interfere with detection of blurred reflection features from
the inner disk.

We propose an alternative explanation for $R_{\rm f} < 1$ based on the
strong dependence of the reflected spectrum on the angle at which an
illuminating photon strikes the disk. This angle crucially determines
the characteristic depth in the disk at which the photon interacts;
this in turn affects the limb-darkening/brightening of the disk
\citep{svo09,gar14a}. A deficiency of {\tt reflionx}, \relxill,
\xillver\ and other widely-used reflection models is the simplifying
assumption of a fixed incidence angle of 45~deg. However, a larger angle
of incidence (measured with respect to the normal to the disk plane),
for example, results in a hotter surface layer and therefore a weaker
reflection signature \citep[see Figure~5 in][]{dau13}.

To test whether the assumption of near-grazing illumination
substantially increases fitted values of $R_{\rm f}$, we produced a new
table of \xillver\ reflection models with a fixed incidence angle of
85~deg and merged them with {\sc relline} to create a new
high-incidence-angle version of \relxill\
(Section~\ref{sec:refspec}). Fitting Spectra A--F as in
Section~\ref{sec:fit1} (i.e., {\sc jf-i}), the fit is slightly worse
($\Delta\chi^2=9.71$) but statistically comparable and still quite
reasonable ($\chi^2_{\nu}=1.09$).  Notably, for the 85-deg model we find
that $R_{\rm f}$ increases with luminosity and that $R_{\rm f}>1$ for
the three most luminous spectra (A--C). Meanwhile, all the other
parameters are consistent with those for {\sc jf-i}
(Table~\ref{tab:fit1}). Importantly, the Fe abundance remains unchanged.

\begin{figure}
\centering
\includegraphics[scale=0.55]{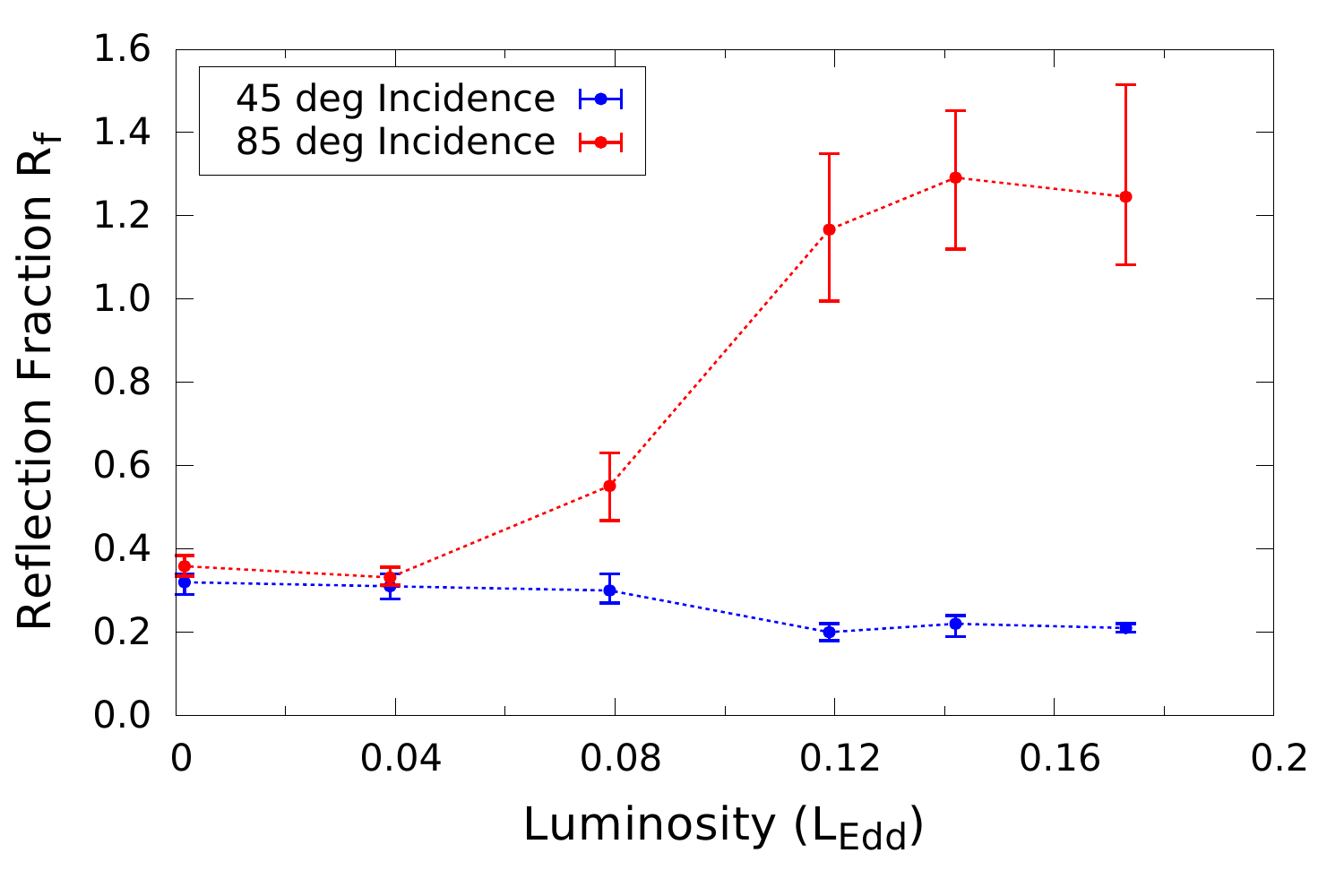}
\caption{
Comparison of the reflection fraction versus luminosity for two
fixed value of the incident angle of the illuminating radiation: the 45
deg value (blue points), which is widely assumed in reflection modeling,
and 85 deg (red points), the value we have assumed for this test.
}
\label{fig:lowi}
\end{figure}

Figure~\ref{fig:lowi} compares the reflection factors computed for the
two models. The large-angle model is more in accord with expectation,
namely, the reflection fraction trends upward with luminosity and the
values at the higher luminosities (with $R_{\rm in}$ near the ISCO and
with correspondingly large reflector area) are sensibly
$\gtrsim1$. Thus, our results qualitatively suggest that the accretion
disk in GX~339--4 is illuminated at near-grazing angles with respect to
the surface of the disk.

Within a few gravitational radii of the horizon, the extreme bending of light
rays causes photons to strike the disk over a wide range of angles
(\citealt{dau13}; see in particular the middle panel of their Figure~5). Given
the strong dependence of $R_{\rm f}$ on the angle of incidence, reaching firm
conclusions concerning the reflection factor will require building a new
generation of models, a task beyond the scope of this paper that will be
addressed in future work.

\subsection{Ionization parameter and geometry}\label{sec:logxi}

As expected, both the normalization $N_{\rm r}$ and the ionization
parameter $\xi$ of the blurred reflection component (\relxill) increase
with luminosity (Table~\ref{tab:fit1}; Figure~\ref{fig:params},
middle-left panel). In particular, the ionization parameter changes from
$\xi=112.2$ to $\xi=2041.7$, which traces very well the ten-fold
increase in luminosity, from 1.6\%~$L_\mathrm{Edd}$ to
17\%~$L_\mathrm{Edd}$.  For the strongly-illuminated portion of the
disk, the variations in $\xi$ and $L$ deviate mildly from the simple
relation $\xi=L/n D^2$, where $n$ is the density of the gas (fixed at
$n=10^{15}$~cm$^{-3}$ for the \relxill\ and \xillver\ models used here
\footnote{The choice of gas density is relatively unimportant; it is the
  ionization parameter that largely determines the properties of the
  reflected spectrum \citep[see][]{gar10,gar13a}.}, and $D$ is the distance from the coronal source to the
strongly heated portion of the disk.  Thus, $D$ increases only modestly
as the luminosity decreases by an order of magnitude in passing from
Spectrum A to Spectrum F:
\begin{equation}
\frac{D_F}{D_A} = \sqrt{\frac{L_F\xi_A}{L_A\xi_F}} \sim 1.3.
\end{equation}
This small change is reasonable given the correspondingly mild increase
in the inner radius obtained for {\sc jf-i}:
\begin{equation}
\frac{R_F}{R_A} = 2.2.
\end{equation}
Likewise, the normalization $N_{\rm x}$ of the unblurred reflection
component (\xillver) increases with luminosity (Figure~\ref{fig:params},
bottom-right panel); although presumably the ionization parameter of
this component also increases with luminosity, we approximate the state
of the gas in the distant reflector as cold and neutral.

%
%
%

\section{Summary and Conclusions}\label{sec:conc}

We have presented an analysis of six composite \rxte\ PCA spectra of the X-ray
binary black hole GX~339--4. All these spectra were taken when the source was
in the hard state. The spectra correspond to luminosities ranging from 1.6\% to
17\% of the Eddington luminosity. The six spectra, each spanning the energy
range 3--45~keV, comprise in total 77 million counts and a total exposure time
of 196~ks. A unique feature of this work is our use of the tool \pcacorr,
which allows us to calibrate the PCA data to a precision of 0.1\%.

The spectra individually, and jointly, are well fitted by a model with three
principal components: \relxill, our model of relativistic ionized reflection;
\xillver, a minor component that models the effects of a cold, distant
reflector; and {\tt Tbabs}, a standard model of Galactic absorption. We
include an ad hoc Gaussian component ({\tt gabs}) to model an absorption
feature near 7~keV.  The origin of this feature is unclear, but it is likely an
artifact resulting from a misestimate of the PCA energy resolution.

We performed two joint fits of the six spectra. In the first of these,
we fixed the spin to its maximal value, which allows the inner-disk
radius $R_\mathrm{in}$ to approach the ISCO radius, and we derived
precise estimates for the evolution of $R_\mathrm{in}$ with
luminosity. We find that the disk becomes increasingly truncated with
decreasing luminosity. Specifically, as the luminosity ranges from 17\%
to 1.6\% of Eddington, $R_\mathrm{in}$ increases from $2.1 R_g$ to $4.6
R_g$. While this trend has been previously reported
\citep[e.g.,][]{pet14, kol14,pla15}, our values of $R_\mathrm{in}$ for
comparable values of luminosity are much smaller than those found by
others. The grossest discrepancy is the hundredfold larger values
reported by \cite{pla15}.

That we find such small values of the inner-disk radius and no evidence for a
thermal disk component is at odds with the current models.  This is particularly
true for Spectrum A with $R_\mathrm{in} = 2.1 R_g$.  One expects
such a modestly-truncated disk to be sufficiently hot \citep[particularly because it
is heated by the corona; e.g.,][]{haa93b} that we should have
detected it with the PCA.  This implies that our model somewhat underestimates
the true value of $R_\mathrm{in}$ and that our model is incomplete. To address
this problem, we are in the process of exploring an extended model that
self-consistently treats the thermal, power-law and reflected components. This
is a challenging problem whose solution is beyond the scope of this paper.

Our analysis indicates that the factor of $\sim 100$ range in the values of
$R_\mathrm{in}$ at fixed luminosity, which have been reported in the
literature, is unlikely to result from the use of different reflection models;
the shifts in $R_\mathrm{in}$ attributable to this cause appear to be
relatively minor.  Instead, the large disparity appears to be attributable to
limitations of the data, one of which is the well-known effects of pileup.  In
this paper, we highlight a particularly important effect, namely, the modest
statistical quality of most data, which has resulted in observers fitting the
blurred reflection component assuming that the Fe abundance is solar, whereas
we demonstrate that super-solar Fe abundance is required for fits to data with
extreme statistical precision. Specifically, we strongly constrain the Fe
abundance (in solar units) to be $A_\mathrm{Fe}=5.0^{+1.2}_{-0.4}$, which is
the average value for our two joint fits.  This strict requirement of the data
is a promising and likely explanation for why, at luminosities $\sim1$\% of
Eddington, we find evidence for relatively mild disk truncation compared to
earlier studies.

We acknowledge that the accuracy of our results are limited,
systematically, by the presence of an absorption feature near 7.2~keV
(Section~\ref{sec:gabs}) whose origin is unknown. However, our
principal conclusions regarding the inner radius of the disk are sound,
being subject to a minor uncertainty of about $20\%$ arising from whether
or not this feature is included in the model.

As the source luminosity and the radiation field bathing the disk grow, the
disk becomes increasingly ionized and its structure changes as $R_\mathrm{in}$
shrinks. At the same time, the large and steady decrease in the high-energy
cutoff indicates that the illuminating coronal source is likewise evolving, as
its temperature drops and its optical depth increases.

In the second of our two joint fits to the six spectra, we made the
standard assumption used in estimating black hole spin, namely, we fixed
$R_\mathrm{in}$ to the radius of the ISCO. Doing so, we constrained the
spin of the black hole to be $a_*=0.95^{+0.03}_{-0.05}$. We were able to
achieve this statistical precision despite the limited spectral
resolution of the PCA because of the quality of the data and its
calibration.

If there is some truncation of the inner disk (i.e., $R_\mathrm{in} >
R_\mathrm{ISCO}$), then the spin is greater than the estimate given
above.  Our estimate of spin agrees well with previous determinations
made using the Fe line method \citep{mil04,mil06,mil08,rei08}.  It is,
however, inconsistent with the upper limit of $a_* <0.9$ derived using
the continuum-fitting method \citep{kol10}, a result that is uncertain
because the accurate values of black hole mass, disk inclination and
distance that are required for successfully applying the
continuum-fitting method are unknown for GX~339--4.  Our result is also
formally incompatible with the value of spin predicted for GX~339--4\ by
\citet{ste13} based on the relationship between spin and jet power
proposed by \cite{nar12}, which has been challenged by \citet{rus13}.

We also obtain a precise estimate for the inclination of the inner disk
of $i=48.1^{+1.0}_{-1.3}$~deg. This value is further subject to an
estimated systematic uncertainty of about $4$~deg arising from whether or
not one chooses to include the 7.2~keV absorption feature in the
model. Our value is inconsistent with the low values found earlier using
the Fe-line method \citep{mil04,mil06,mil08,rei08}, while it is more in
line with reasonable expectations for the mass of the black hole based
on the value of the mass function.

%
%
%
%
\acknowledgments  We thank Felix F\"{u}rst, Mike Nowak, Tim Kallman, Rubens
Reis, Francesco Tombesi, and Andrzej Zdziarski for useful and valuable
discussions. JG and JEM acknowledge the support of NASA grant NNX11AD08G. JFS
has been supported by NASA Hubble Fellowship grant HST-HF-51315.01. VG
acknowledges support provided by NASA through the Smithsonian Astrophysical
Observatory (SAO) contract SV3-73016 to MIT for support of the Chandra X-Ray
Center (CXC) and Science Instruments; CXC is operated by SAO for and on behalf
of NASA under contract NAS8-03060. 

{\it Facility:} \facility{RXTE}
%
%
%
%
\bibliographystyle{apj}
\bibliography{my-references}
%
%
%
%
\end{document}